\documentclass[article]{JHEP3}

\usepackage{amsmath,epsfig}

\usepackage{amssymb,amsfonts}

\title{Open strings in relativistic ion traps}
\preprint{\hepth{0302159}\\LPTHE-03-08}
\author{B. Durin and B. Pioline\\
LPTHE, Universit\'es Paris 6 et 7, 4 place Jussieu, \\
75252 Paris cedex 05, FRANCE\\{\tt E-mail:
bdurin@lpthe.jussieu.fr, pioline@lpthe.jussieu.fr}}
\abstract{Electromagnetic plane waves provide examples of time-dependent
open string backgrounds free of $\alpha'$ corrections. 
The solvable case of open strings in a quadrupolar wave front, 
analogous to pp-waves for closed strings, is discussed. In light-cone
gauge, it leads to non-conformal boundary conditions similar to 
those induced by tachyon condensates. 
A maximum electric gradient is found, at which macroscopic strings
with vanishing tension are pair-produced -- a non-relativistic analogue 
of the Born-Infeld critical electric field.
Kinetic instabilities of quadrupolar electric fields are cured by 
standard atomic physics techniques, and do not interfere with the
former dynamic instability. A new example of non-conformal open-closed
duality is found. Propagation of open strings in
time-dependent wave fronts is discussed. }

\newenvironment{theorem}{\begin{quote} \footnotesize }{ \normalsize \end{quote}}

\newcommand{\p}{\partial}

\newcommand{\nn}{\nonumber}

\newcommand{\Real}{\mathbb{R}}
\newcommand{\Zint}{\mathbb{Z}}

\newcommand{\Tr}{\mbox{Tr}}

\def\bea{\begin{eqnarray}}
\def\eea{\end{eqnarray}}
\def\be{\begin{equation}}
\def\ee{\end{equation}}
\def\ba{\begin{align}}
\def\ea{\end{align}}
\def\bse{\begin{subequations}}
\def\ese{\end{subequations}}
\def\bi{\begin{itemize}}
\def\ei{\end{itemize}}

\def\e{\epsilon}

\def\om{\omega}

\newcommand{\ps}{\partial_\sigma}
\newcommand{\pt}{\partial_\tau}

\newcommand{\rhs}{{\sc rhs }}
\newcommand{\lhs}{{\sc lhs }}
\newcommand{\tr}{\text{tr }}
\newcommand{\epn}{\, e}

\begin{document}
\maketitle
\tableofcontents
\section{Introduction}

While huge efforts have been devoted to finding phenomenologically viable
static backgrounds for string theory, the study of time-dependent
string backgrounds has only started to receive due attention recently.
While a string implementation of inflation is still lacking, several
examples of cosmological singularities have been studied 
\cite{Horowitz:ap,Nekrasov:2002kf,lms, Elitzur:2002rt, Craps:2002ii,
Cornalba:2002nv}, in first quantized string theory,
raising strong doubts as to the validity of string perturbation techniques.
Regardless of the issue of singularities, our understanding
of string theory in time-dependent backgrounds is still very limited,
and can only benefit from the study of solvable, if simplistic, cases.

Gravitational plane waves are well-known examples of time-dependent
backgrounds, free of string corrections \cite{Amati:1988ww,hs,Tseytlin:1992pq,
Nappi:1993ie}. They have received renewed interest 
recently in relation to the AdS/CFT correspondence,
as they appear in a particular limit of AdS and other backgrounds
\cite{Blau:2001ne}. They can be studied both 
at the first quantized level and in string
field theory using light-cone techniques, even in the presence of
Ramond backgrounds \cite{metsaev}. After fixing the light-cone
gauge, the worldsheet theory acquires a mass gap proportional to
the light-cone momentum $p^+$, so that high energies on
the worldsheet correspond as usual to long wavelength in target space.
In fact, non-conformal two-dimensional sigma models can be turned into
on-shell backgrounds by dressing them appropriately
with dilaton gradients and fluxes in the light-cone directions \cite{mm}.
Time-dependent gravitational plane-waves have been recently studied in
\cite{Gimon:2002sf, Papadopoulos:2002bg, Blau:2002js}.

Less familiar perhaps is the fact that electromagnetic waves can
be exact open string backgrounds, free of $\alpha'$ corrections.
This was first recognized in the context of the Born-Infeld string
 \cite{thorlacius},
where it was shown that a transverse displacement of a D$p$-brane
$\Phi(x_i)$, when supplemented by an electric field $F_{0i}=\p_i \Phi$,
is an exact open string background if $\Phi$ is an harmonic
function of $x_i$. 
Upon thinking of $\Phi$ as the extra component
of an higher dimensional gauge field $A_{p+1}$, this amounts to considering a
configuration with null electromagnetic field strength $F=\p_i \Phi dx^i
\wedge (dx^0-dx^p)$, i.e. an electromagnetic plane wave. 
The case  of a uniform, time-dependent $F_{i+}$, or rather its T-dual version, 
was considered recently in \cite{bh,Bachas}. It includes the 
case of a monochromatic plane wave discussed long ago in 
\cite{Ademollo}. In this
paper, we consider the generalization of these exact string
backgrounds to a
time-dependent non-uniform plane wave accompanied by a constant and uniform 
magnetic field in the transverse direction,
\be
\label{wav}
F= \p_i \Phi(x^+,x^i)~ dx^i\wedge dx^+ ~+~\frac12 B_{ij} ~dx^{i}\wedge
dx^j\ .
\ee
Here $\Phi$ is an harmonic function in transverse space, for the open
string metric derived from $B$ \cite{Abouelsaood:gd,sw}, 
$G_{ij}=\delta_{ij}+(2\pi \alpha')^2 B_{ik}B_{jk}$. 
In the light-cone gauge, this breaks
conformal invariance through the boundary condition
\be
\p_\sigma X^i + p^+ \p_i\Phi (X^i,X^+) + B_{ij} \p_\tau X^j = 0\ ,\qquad
\sigma\in \{0,\pi\}  \label{boundaryconditions}
\ee
while preserving conformal invariance in the bulk, 
$(\p_\tau^2-\p_\sigma^2)X^i=0$. For general $\Phi$, this implies a complicated
non-linear and non-local coupling between the left and right movers. 
This problem however becomes linear for $\Phi$ quadratic in $x^i$, 
corresponding 
to a quadrupolar electric potential. Our main goal in this paper will be 
to solve for the dynamics of first quantized open strings in this
background. More generally, we
shall consider open strings ending on two D-branes with
possibly different quadrupolar electric potential and uniform magnetic
fields. 

The non-conformal boundary deformation \eqref{noconf} 
is very reminiscent of deformations that arose in studies of tachyon
condensation and background independent string 
field theory \cite{wbsft,Shatashvili:1993kk,Bardakci:2001ck,Arutyunov:2001nz}, 
with two important caveats: (i) the deformation is taken on the (null)
boundary of the strip, rather than the disk or the annulus
and (ii) the ``tachyon 
field" $\Phi$ satisfies $\Delta \Phi=0$ in contrast to 
$\Delta T+2 T=0$. In particular, the potential $\Phi$ is unbounded
from above and from below and has no global minimum. This implies that,
in the absence of a magnetic field, charged particles and 
open strings will be repelled away
from the center of the quadrupolar potential.  This is the direct analogue
of the defocusing of geodesics by a purely gravitational plane 
wave. This kinetic instability does not mean that the background
is unstable, on the contrary 
studies of the gravitational wave case have concluded
to their stability \cite{Brecher:2002bw}.

Let us outline our main findings: 
\begin{enumerate}
\item[i)] For a critical value of the electric gradient, 
we find an instability towards producing
macroscopic strings at no cost of energy: for that value, the
tensive energy of an open string, which goes like its length square,
is overwhelmed by the energy gained by stretching the string in
the quadratic potential. The string then grows arbitrarily until it
reaches the condensator that created the quadrupolar
field in the first place and discharges it. This 
critical {\it electric gradient}
is a non-relativistic analogue of the critical {\it electric field strength}
in ordinary Born-Infeld electrodynamics. A simple
elastic dipole model suffices to capture it.

\item[ii)] In the absence of a magnetic field, the dynamical instability
towards producing macroscopic strings is shadowed by the 
kinetic instability of charged particles  in a quadrupolar electrostatic 
potential. One way to cure the latter is to 
switch on a constant magnetic field in the unstable directions,
thereby confining the string into Larmor orbits. This is the
familiar Penning trap configuration used in ion trapping studies. 
We show that the critical gradient instability is impervious to this change.

\item[iii)] Another way to achieve confinement of charges in a quadrupolar
electric field is to modulate the electric field at a resonant frequency
with the proper modes of the electrostatic configuration: this is known as 
the Paul trap or rf (radiofrequency trap). We show that it manages equally
well to confine charged open strings. We further compute the effect
of a non-resonant adiabatic variation of the electric field.

\item[iv)] While the one-loop open string amplitude is trivial due to the null
nature of the electromagnetic field, this is no longer the case when
the light-cone coordinate $x^+$ is compactified. Under this assumption,
we compute the one-loop free energy for open bosonic
strings, and find that it coincides with the overlap of boundary states
in the closed string channel. This gives a new example of open-closed duality
in the non-conformal regime, beyond the ones found 
in \cite{Bergman:2002hv,Gaberdiel:2002hh}. 

\item[v)] We consider the propagation of open strings in a
 time-dependent quadrupolar electric potential. As usual for 
plane wave background there is no production of strings at
zero coupling \cite{Aichelburg}, however strings do get excited as they cross
the electromagnetic wave. We compute the mode production in 
Born's approximation, and find as in \cite{hs} that impulsive wave profiles,
unlike shock waves, transfer an infinite amount of
energy into the string. We observe that the Bogolioubov matrix fails to
incorporate an hidden degree of freedom of the background, 
for which the backreaction can be computed straightfowardly.
We also discuss the adiabatic and sudden approximations. 
\end{enumerate}

These results call for a number of further questions: $\bullet$ 
first and foremost, in analogy with the NCOS construction \cite{min} 
can one take a scaling limit towards the critical electric
gradient, and decouple
the quasi-tensionless macroscopic open strings from the closed strings ?
$\bullet$ Can a large $N$ limit of such D-brane
configurations give rise to an holographic dual to gauge theories with
time-dependent light-like 
non-commutativity \cite{ Aharony:2000gz,Dolan:2002px} ?
$\bullet$ Can one take into account in a
consistent way the condensation of these macroscopic strings ? More generally,
can one take into account the backreaction of open strings 
with $p^+ \neq 0$ onto the background \eqref{wav} self-consistently ?
Is the backreaction finite or does it destroy the background
irremediably ? 
Can one compute string production at finite string coupling ? 
$\bullet$ Can one generalize this construction to other integrable
boundary deformations, such as higher derivative Gaussian operators
$\oint X \p^n X$, corresponding to excited string states, 
or Sine-Gordon-type boundary deformations ?
$\bullet$ Can one understand further the relation between worldsheet
RG flow and target space amplitudes ? What is the space-time interpretation of
a boundary RG flow between two conformal fixed points ?
$\bullet$ What are the implications of the critical electric gradient
for the higher-derivative corrections to the Born-Infeld action ?
$\bullet$ It has been observed \cite{Nekrasov:2002kf,bh} that open
strings in electric or null electric 
fields share many features with twisted closed
strings at a spacelike or null orbifold 
singularity \cite{Nekrasov:2002kf,lms}. Is there
a gravitational analogue to this critical gradient instability ?
$\bullet$  Finally, at a far more futuristic level, one may wonder
whether Bose-Einstein condensates of strings may be obtained
using our string traps, or if the critical gradient instability 
of charged elastic dipoles may be one day be observed in QCD.
We hope to come back to some of these questions in the future.

The plan of the paper is as follows. In Section 2, we show that the 
configuration \eqref{wav} gives an exact open string background,
identify the solvable cases, discuss the motion of charged particles
in this background, and how the transverse motion can be stabilized.
In Section 3, we introduce an elastic dipole model and find a
critical electric gradient at which it becomes tensionless. We show
how adding a magnetic field or modulating the electric field stabilizes
the kinetic instability, while preserving the phase transition. 
In Section 4, we carry out the first quantization of open strings in
a quadrupolar electric potential, possibly stabilized by a magnetic
field. The one-loop amplitude is computed and open-closed
duality is checked. 
Time dependent electromagnetic waves are considered in 
Section 5, in the Born, adiabatic and sudden approximations. 
Our conclusions are contained in the introduction.
A mathematically rigorous derivation of the stability diagram can be
found in Appendix A, a derivation of the modular property of the
string partition functions is provided in Appendix B,
and somewhat unwieldy normalization factors are gathered in Appendix C.

\section{Exact electromagnetic waves and ion traps}
Null electromagnetic fields $F=F_{i+} dx^i \wedge dx^+$ are particularly simple
as they provide exactly conformal deformations of the open string
boundary conditions. Here we generalize the solution of \cite{thorlacius}
to include time dependence and a constant magnetic field\footnote{The
possibility to add a magnetic field while preserving exactness was
observed in a discussion with A. Tseytlin.}, and discuss
how the light-cone gauge breaks conformal invariance on the boundary.
We further describe the motion of charged particles in such backgrounds,
and how their kinematical instability may be cured.

\subsection{Electromagnetic waves as exact open string backgrounds}
\label{generalsetting}
Let us consider the Abelian gauge field
\be
\label{ab}
A= \Phi(x^i,x^+) dx^+ + B_{ij} x^i dx^j\ ,\quad F=dA\ ,
\ee
where $\Phi(x^i,x^+)$ is an arbitrary function independent of $x^-$,
and $B_{ij}$ a constant antisymmetric matrix. One may show
that $A$ solves the equations of motion of Born-Infeld electrodynamics,
\be
\p_\mu \left[ (\det G_{\alpha\beta})^{1/4} G^{\mu\rho} F_{\rho\nu} \right]=0
\ee
where $G_{\mu\nu}=(G^{\mu\nu})^{-1}=\eta_{\mu\nu}+(2\pi \alpha')^2
F_{\mu\sigma} \eta^{\sigma\rho}F_{\nu\rho}$ is the open string
metric \cite{Abouelsaood:gd,sw},
\be
\label{osm}
ds^2 = 2 dx^+ dx^- + G_{ij} dx^i dx^j + 
(2\pi\alpha')^2 
\left[ \left| \p_i \Phi(x^+,x^i) \right|^2 
dx^+ +  \p_i \Phi(x^+,x^i) B_{ij} dx^j  \right] dx^+ 
\ee
under the only condition that $\Phi$ be an harmonic function in transverse
space,
\be
\label{harm}
\frac{1}{\sqrt{G}} \p_i \left( \sqrt{G} ~ G^{ij}  \p_j \right)
\Phi(x^i, x^+)=0\ .
\ee
In this equation, the volume element does in fact drop since the
transverse part $G_{ij}$ of the open string metric depends only
on the constant magnetic field, $G_{ij}=\delta_{ij}+
(2\pi \alpha')^2 B_{ik} B_{jk}$. For two transverse directions, $G_{ij}$
is proportional to the identity matrix, so \eqref{harm} reduces
to $\p_i^2 \Phi=0$, hence $\Phi$ has to be the real part of
an holomorphic function. It is interesting to note that the open
string metric \eqref{osm} is a gravitational plane wave in 
Brinkmann coordinates, together with an extra magnetic coupling.
Note also that the energy momentum tensor associated 
to the configuration \eqref{wav} is given by
\be
T^{\mu\nu}=\frac12 \sqrt{
-\det(\eta_{\alpha\beta}+2\pi \alpha' F_{\alpha\beta})} G^{\mu\nu}
-\frac12 \eta^{\mu\nu}
\ee
where $G^{\mu\nu}$ is the inverse of the open string metric appearing
in \eqref{osm}, and we subtracted the rest energy of the D-brane 
itself. The light-cone momentum and energy 
$P^{\pm} = \int dx^i dx^- T^{\pm+}$ therefore vanish irrespective
of the profile $\Phi(x^i,x^+)$. Furthermore, one may check that
the background \eqref{ab} preserves half of the supersymmetries,
namely those satisfying $\Gamma^+ \epsilon=0$.

In addition to solving the Born-Infeld equations of motion, one can argue
that the background \eqref{ab} subject to condition \eqref{harm} is
in fact an exact solution of open string theory to all orders in $\alpha'$.
Indeed, as the gauge field in \eqref{ab} is independent of $x^-$, the
embedding coordinate $X^+$
cannot be Wick contracted and remains classical. The sole effect of the
magnetic term in the boundary deformation $\oint A_\mu dX^\mu$
is to renormalize the transverse metric from $\delta_{ij}$ to $G_{ij}$.
The same reasoning as in \cite{thorlacius} then guarantees that there
are no divergences coming from the electric term $\oint \Phi(x^i,x^+)
dx^+$ as long as $\Phi$ is harmonic in transverse space. 

The Abelian configuration \eqref{ab} may be generalized further to the 
non-Abelian $U(N)$ case, by choosing a different electric field $\Phi$ 
for each generator of the gauge group: as only $A_+$ is non vanishing, 
the non-Abelian part
of the field strength $F=dA+[A,A]$ remains null as it should. The
magnetic field $B_{ij}$ should remain Abelian however.

\subsection{Light cone gauge and non-conformal boundary conditions}

In the presence of the $\oint A_\mu dX^\mu$ boundary deformation,
the embedding coordinates of the string $X^i(\sigma,\tau)$
are therefore still free fields on the worldsheet,
\be
\label{2dlap}
(\p_\tau^2 - \p_\sigma^2) X^\mu=0\ ,\quad
\ee
but with the non standard boundary conditions (setting $2\pi\alpha'=1$)
\begin{subequations}
\label{bc}
\bea
\p_\sigma X^+&=&0  \\
\p_\sigma X^i+ \p_i \Phi_a(X^+,X^i) \p_\tau X^+ + B_{ij} \p_\tau X^j 
&=& 0   \qquad\mbox{at} ~~
\sigma=\sigma_a\ ,\quad \label{bci}\\
\p_\sigma X^- - \p_i \Phi_a(X^+,X^i) ~ \p_\tau X^i&=& 0 
\eea
\end{subequations}
where we assumed different electric potentials,  $\Phi_0$ and 
$\Phi_1$, at the two ends of the string, $\sigma_0=0$ and $\sigma_1=\pi$,
as in the case of a general non-Abelian configuration.
By \eqref{2dlap} $X^i$ can be decomposed in left and right movers:
\be 
X^i(\tau, \sigma) = f^i(\tau + \sigma) + g^i(\tau - \sigma) \label{lrmovers}
\ee
The conditions \eqref{bc} are compatible with choosing the light-cone gauge
$X^+=x_0^+ + p^+ \tau$, upon which the condition \eqref{bci} becomes
\be
\p_\sigma X^i+ p^+ \p_i \Phi_a(X^+,X^i)+ B_{ij} \p_\tau X^j 
= 0   \qquad\mbox{at} ~~
\sigma=\sigma_a\ ,\quad \label{bcilc}
\ee
corresponding to a  boundary perturbation:
\be \int d\tau ~ 
\left( p^+ \Phi(X^+,X^i) + \frac12 B_{ij} X^j \frac{dX^j}{d\tau} \right) 
\label{noconf}.\ee 
As usual, choosing the light-cone gauge breaks conformal invariance, 
but unlike gravitational waves, electromagnetic waves preserve
conformal invariance in the bulk of the two-dimensional field theory. 
On the other hand,
such non-conformal boundary deformations are reminiscent of the
tachyonic perturbations considered in the context of background
independent string field theory \cite{wbsft}, with two important caveats:
(i) since $\Phi$ is harmonic, it cannot be bounded
from below or above, in contrast to the tachyonic perturbations
usually considered; (ii) the deformation is taken on the boundary
of the strip, and is inequivalent to a deformation
$\oint d\tau \Phi(x^i)$ on the boundary of the disk since
conformal invariance is broken. Nevertheless, it is interesting
that electromagnetic plane waves can provide a real-time setting
to study non-conformal boundary deformations.

\subsection{Solvable null electric fields and T-duality}
While the equation \eqref{bcilc} in general implies an inextricable
non-local relation between the left and right-moving components
\eqref{lrmovers} of the field $X^i$.
we will be interested in cases where this relation becomes solvable.
This is in particular the case when 
\begin{enumerate}
\item[1)] $\Phi_a=(A_a)_i (x^+) x^i$ is linear 
in $x^i$,
yet an arbitrary function of $x^+$: this corresponds to an electric
field $F_a=(F_a)_i(x^+) dx^i dx^+$ uniform in the transverse coordinates. The
equation  \eqref{bcilc} becomes a first order equation with source,
\be
\p_\sigma X^i + B_{ij}
\p_\tau X^j + p^+ (F_a)_i (x^+) = 0   \qquad\mbox{at} ~~
\sigma=\sigma_a\ .\quad \label{bclin}
\ee
Conformal invariance is therefore unbroken, and the field $X^i$ can
be computed easily by linear response. 
For $B=0$ this case is T-dual to a configuration of two D0-branes 
following null trajectories
$Y^i_a(x^+)$ where $dY^i_a/dx^+ = - A^i_a (x^+)$. Since this case
has been considered recently in \cite{Bachas}, 
we will not discuss it any further.

\item[2)]  $\Phi_a=(h_a)_{ij} (x^+) x^i x^j/2$   is quadratic  
in $x^i$, and arbitrary function of $x^+$: this corresponds to an electric
field $F_a=(h_a)_{ij}(x^+) x^i dx^j dx^+$ which now has a constant 
gradient in the transverse coordinates $x^i$. The
equation \eqref{bcilc} is still linear but includes a mass contribution,
in general time-dependent:
\be
\p_\sigma X^i + p^+ (h_a)_{ij} (x^+) X^j + \frac12 B_{ij} \p_\tau X^j
= 0   \qquad\mbox{at} ~~
\sigma=\sigma_a\ .\quad \label{bcquad}
\ee
Conformal invariance is now genuinely broken. This will be the case
of most interest in this paper, where we will show that the system
exhibits an interesting phase transition at a critical value of the
gradients $(h_a)_{ij}$.

\item[3)] Clearly, the two cases above can occur simultaneously,
for  $\Phi_a=(A_a)_i (x^+) x^i + (h_a)_{ij} (x^+) x^i x^j/2$. This amounts
to varying in time the location of the saddle point 
of the quadratic field, $x^i_{sad} = - (h^{-1})^{ij} A_j$, and again
could be computed by linear response from \eqref{bcquad}.
\end{enumerate}
It would be interesting to investigate the behaviour of open strings
for null electric fields with a critical point of higher order, but
this will not be addressed in this work.

Let us pause to discuss the T-dual version of the boundary deformation
\eqref{bcquad}. Since the boson $Xî$ are free fields in the bulk, they
may be dualized into free bosons 
$\tilde X^i(\tau,\sigma)=f^i(\tau+\sigma)-g^i(\tau-\sigma)$. 
In terms of $\tilde X^i$, the non-conformal boundary condition
\eqref{bcquad} becomes, after differentiating with respect to $\tau$
(and assuming $h_a$ to be constant),
\be 
\p^2_\tau \tilde{X}^i + p^+ (h_a)_{ij} \ps \tilde X^j = 0,  \label{bounddef}
\ee
This corresponds to a boundary deformation $\int (h^{-1})_{ij} 
\p_\tau X^i \p_\tau X^j d\tau/p^+$ by a massive operator of the open string
spectrum. Physically, it amounts to attaching a (possibly tachyonic) mass
$h^{-1}/p^+$ at the ends of the string. Notice that deformations
by higher massive string states $\int m_{ij}^{(n)} \p_\tau X^i \p_\tau^n X^j
d\tau $ also give rise to integrable boundary conditions.
This may be useful in evaluating the effective action for massive
string excitations. On the other hand, we may consider another variation on our
boundary problem \eqref{bcquad}, where $\tau$ and $\sigma$ derivatives
are exchanged:
\be
\label{notdual}
\p_\tau X^i + p^+ (h_a)_{ij} (x^+) X^j - \frac12 B_{ij} \p_\sigma X^j
= 0   \qquad\mbox{at} ~~
\sigma=\sigma_a\ .\quad 
\ee
Despite apparences, this boundary deformation 
is {\it not} T-dual to \eqref{bcquad}, although it is still
integrable by the same techniques. Instead, for vanishing $B$
one may integrate \eqref{notdual} with respect to $\tau$,
and find that it describes open strings ending on infinitely boosted D-branes 
falling along the field lines of $\Phi$.

\subsection{Charged particle in a null electric field}
As a warm-up, let us consider a relativistic particle of mass $m$
and unit charge in an arbitrary null electric field $A= \Phi(x^+,x^i) dx^+$
satisfying the harmonicity constraint \eqref{harm}, together with
a constant magnetic field. The equations of motion following from the action 
\be
\label{act}
S=\int \left[ \frac{1}{2e} \left( \p_\tau X ^\mu \right)^2 - e ~ m^2 \right]
d\tau + A_\mu dX^\mu
\ee
read, after choosing the gauge $e=1$, 
\be
\begin{aligned}
(d^2/d\tau^2)& X^+ = 0 \\
(d^2/d\tau^2)& X^i + \p_i \Phi~ \p_\tau X^+ + B_{ij} \p_\tau X^j =0 \\
(d^2/d\tau^2)& X^- - \p_i \Phi~ \p_\tau X^i =0
\end{aligned}
\ee
The first equation can be integrated to $X^+(\tau)=x_0^+ + p^+ \tau$.
Taken together, the second and third equations imply that
\be
\label{msp}
H= \frac12 (p_i)^2 + p^+ p^- + p^+ \Phi(X^+,X^i) + m^2
\ee
is a constant, where $p_i=\p_\tau X^i- B_{ij} X^j$ 
and $p^-=dX^-/d\tau- \Phi$ are 
the canonical momentum conjugate to $X^i$ and to $X^+$. It is in fact
the Hamiltonian constraint\footnote{A dimensionally correct Hamiltonian
would be obtained by choosing $e=1/m$ instead, leading to $H\to H/m$.}
enforced by the Lagrange multiplier $e$
in \eqref{act}, hence the constant can be set to 0, thereby enforcing
the mass shell condition. The motion thus becomes that of  
a non-relativistic particle in a time-dependent ``electrostatic potential''
$\Phi(x^+,x^i)$. The occurence of non-relativistic kinematics is a standard
feature of light-cone quantization.

\subsection{Kinetic instability and trapping}
\label{iontraps}
Now let us assume momentarily that no magnetic field is present. Since
the potential $\Phi$ is harmonic, there are no local minima, but only
saddle points, hence the motion is expected to be unstable. This
statement should actually be qualified: in two dimensions, a critical
point of the potential will look as $\Phi=\Re(z^n/n)$ with $z=x+i y$. Ordinary
saddle points correspond to $n=2$, $\Phi=\frac12(x^2-y^2)$ and indeed 
the motion along $y$ is that of an inverted harmonic oscillator. 
For higher order critical points however, the motion is easily integrated
to $z(\tau)=[n(n-2)\tau]^{1/(2-n)}$ and therefore converges to $z=0$ for 
late times. We will not be able to solve this non-linear equation 
for the string for $n>2$, hence we will not discuss it further.
On the other hand, there are several known ways to cure the instability
of a quadrupolar electrostatic potential:

\begin{enumerate}
\item[a)] {\it Penning trap:}
A well known scheme used in atomic physics to trap ions is to
add a constant magnetical field to confine the unstable directions
of the electrostatic potential.
The simplest example involves 3 dimensions, with a cylindrically 
symmetric quadrupole potential $V(x)=-\frac{e}{2} (x^2 + y^2 - 2 z^2)$,
and uses a constant magnetic field along $z$ to confine ions in the
$(x,y)$ plane. The equation of motion
\be
\begin{pmatrix} \ddot{x} \\ \ddot{y} \\ \ddot{z} \end{pmatrix}
+ 2b  \begin{pmatrix} -\dot{y} \\ \dot{x} \\ 0 \end{pmatrix}
- e \begin{pmatrix} x \\ y \\ -2z \end{pmatrix}
= \begin{pmatrix} 0 \\ 0 \\ 0 \end{pmatrix}
\ee
leads to the dispersion relation
\be
(\om^2-2 \om b+e)
(\om^2+2 \om b+e)
(2e-\om^2)=0 
\ee
where the 3 terms correspond to the left and right polarizations in 
the $(x,y)$ plane and the $z$ direction, respectively. The motion
is stable iff all roots are real, i.e. $b^2>e$ and $e>0$. Hence only
positively charged (say) particles are trapped, and only if the magnetic field
is sufficiently strong.

\item[b)] {\it Paul trap:}
Charged particles can also be trapped with a quadrupolar field
alone, if it is modulated at a particular frequency,
$\Phi(x^+)=(e + f \sin(\omega \tau)(x^2-y^2)/2$, (parametrically)
resonant with the
proper mode of the quadrupolar field: this is known in atomic physics 
as the Paul, or rf (radiofrequency) trap. In mathematical terms, the
equation describing motion along the $x$ direction (say) is the
Mathieu equation,
\be
\ddot{x} + \left[ e + f \sin(\omega \tau) \right] x = 0
\ee
Its  stability diagram, discussed in \cite{vdpol}, 
is qualitatively the same as for a rectangular 
modulation, which can be computed much more easily and is shown
on Figure \ref{stability}. One sees that for $e<0$, there is a small
range of $f$ where the motion is still stable. We will return to this
point in Section 3.4.

\item[c)] {\it Quadrupolar trap}: Finally, 
while a static quadrupolar electric field 
alone cannot trap neutral particles, it can trap polarizable molecules
with negative polarizability: the potential energy of an induced
electric dipole $\mu_i=\alpha E_i$ is $W=-\alpha E^2$; if $\alpha$
is negative, this is minimal at the center of the quadrupole, where
$E_i$ vanishes \cite{wing}. While the ground state of a molecule has usually positive
polarizability, excited states if degenerate at $E=0$ can have negative 
$\alpha$: the Stark energy shift due to the electric field is given 
by first order degenerate perturbation theory, and is negative for
part of the multiplet. We will not pursue this case here.

\end{enumerate}
These schemes will prove convenient below to construct stable open string
backgrounds. Note however that the instability encountered here is 
rather innocuous,
as it is an instability of the motion of open strings rather than of the
background itself. We will refer to it as a kinematic instability.
Studies of a very similar issue in the context of
gravitational plane waves with a non-positive quadratic form have
found no hint of genuine instability \cite{Brecher:2002bw}.

\section{Elastic dipole in a quadrupolar electric field}
As we have just seen, relativistic
particles of unit charge and light-cone momentum
$p^+$ in the electromagnetic wave \eqref{wav} behave as non-relativistic 
particles of charge $p^+$ in the (in general time-dependent)
electrostatic potential $\Phi$ and uniform magnetic field $B_{ij}$.
In this section, we show that an analogous statement is true for a 
relativistic open string in \eqref{wav}. We consider a simple model
of an elastic dipole in an electrostatic potential, which captures
much of the physics of the string at low energies.


\subsection{Light-cone energetics and elastic dipole model}
Let us first call some basic features of light-cone quantization.
As usual, the equations of motion \eqref{2dlap} and
\eqref{bc} must be supplemented by the Virasoro conditions, 
which are impervious to the electromagnetic field,
\begin{subequations}
\bea
p^+ \p_\sigma X^- &+& \p_\tau X^i \p_\sigma X^i = 0 \label{vir1} \\ 
- p^+ \p_\tau X^- &+& \frac12 \left[ (\p_\tau X^i)^2 + (\p_\sigma X^i)^2 \right]
=0 \label{vir2}
\eea
\end{subequations}
Equation \eqref{vir1} can be used to eliminate $X^-$ in terms of the
transverse coordinates, up to translational zero-modes $x^-_0+ p^-_0\tau$.
On the other hand, the Hamiltonian constraint \eqref{vir2} can be rewritten
in terms of the canonical momenta
\be 
\pi_- = -\p_\tau X^+ = -p^+\ ,\quad \pi_i = \p_\tau X^i\ , \quad
\pi_+=-\p_\tau X^- + \Phi_0( X(0)) \delta_0 - \Phi_1( X(\pi)) \delta_\pi, 
\ee
and integrated from $0$ to $\pi$ to yield the mass-shell condition,
\be
\label{lchamil}
p^- = {\cal H}_{lc} = \frac{1}{2 p^+} \left(\int_0^\pi  \left[
( \pi_i)^2 + ( \p_\sigma X^i)^2 \right] d\sigma \right)~-~ 
 \Phi_0(X(0)) + \Phi_1(X(\pi))
\ee
It is conserved if the potentials $\Phi_a$ are independent of $x^+$.
Comparing \eqref{lchamil} to \eqref{msp}, we see that
the light cone energy can be interpreted as the  
Hamiltonian of two  non-relativistic particles immersed in the 
time-dependent ``electrostatic'' potential
$p^+ \Phi_1(X(\pi)) - p^+ \Phi_0(X(0))$, and bound by an elastic
potential  $\int_0^\pi \frac12 ( \p_\sigma X^i)^2 d\sigma$. It will 
be crucial that the binding energy grows like the length {\it squared}
of the string, in contrast to a linear potential in the usual
relativistic case. 

If we are interested with the low-energy aspects of the propagation
of open strings in \eqref{wav}, we may therefore content ourselves
with a simplified model of a pair of charges, with positions $x_L^i(\tau)$
and $x_R^i(\tau)$, bound by a linearly stretched string,
\be
\label{sstr}
X^i(\tau,\sigma) = \frac{1}{\pi} 
\left[ \sigma ~x_R^i + (\pi - \sigma)~ x_L^i \right]
\ee
The light-cone energy of this ``dipole'' (it is a {\it bona fide} dipole
only in the neutral case  $\Phi_0(x^+,x^i)=\Phi_1(x^+,x^i)$) is simply
\be
\label{ms2}
-p^+ p^- + \frac12 (p_L^2 + p_R^2) + V(x_L,x_R) \equiv 0
\ee
where $V$ is the potential energy
\be
\label{pote}
V(x_L,x_R)= \frac{1}{2\pi}
(x_L^i-x_R^i)^2 + p^+ \left[ \Phi_1(x_R) - \Phi_0(x_L) \right]
\ee
Note that a {\it static} solution of this dipole problem will be an exact 
solution of the string, since \eqref{sstr} will then satisfy the bulk 
equation of motion $(\p_\tau^2-\p_\sigma^2)X_i=0$ and the boundary
conditions
\begin{subequations}
\label{cbc}
\bea
x_R^i-x_L^i ~-~  \pi p^+~\p_i \Phi_0(x_L)  &=& 0   \qquad\mbox{at} ~~
\sigma=0\\
x_R^i-x_L^i ~+~  \pi p^+~\p_i \Phi_1(x_R)  &=& 0   \qquad\mbox{at} ~~
\sigma=\pi
\eea
\end{subequations}
In contrast to the single particle case, the two-body potential \eqref{pote}
is no more harmonic, but satisfies
\be
\left( \Delta_L + \Delta_R \right) V = \frac{2D}{\pi}
\ee
where $D$ is the number of transverse spatial dimensions.
It may therefore admit stable extrema instead of only saddle points.

\subsection{Critical electric gradient and macroscopic strings}
We now consider the dynamics of an elastic dipole in a quadrupolar
electrostatic field, $\Phi_a=(h_a)_{ij} x^i x^j/2$.
In the case where $h_0$ and $h_1$ commute, we can diagonalize them simultaneously 
and the potential energy \eqref{ms2} in a proper direction is then given
by the quadratic form,
\be
V(x_L, x_R) =\sum_{i=k,l} \frac{1}{2\pi}
\begin{pmatrix}x_L & x_R \end{pmatrix}\cdot Q \cdot
\begin{pmatrix}x_L \\ x_R \end{pmatrix}\ ,\quad
Q=\begin{pmatrix} 1-\pi p^+ h_0  & -1 \\ -1 & 1+\pi p^+ h_1
\end{pmatrix} 
\ee
with in general an isolated degenerate point at the origin $x_L=x_R=0$.
It will be convenient to introduce the rescaled electric gradients
$e_a$ defined by:
\be
\label{convention}
e_a = \pi p^+ h_a
\ee

\FIGURE{\label{phasediag}
\hfill\epsfig{file=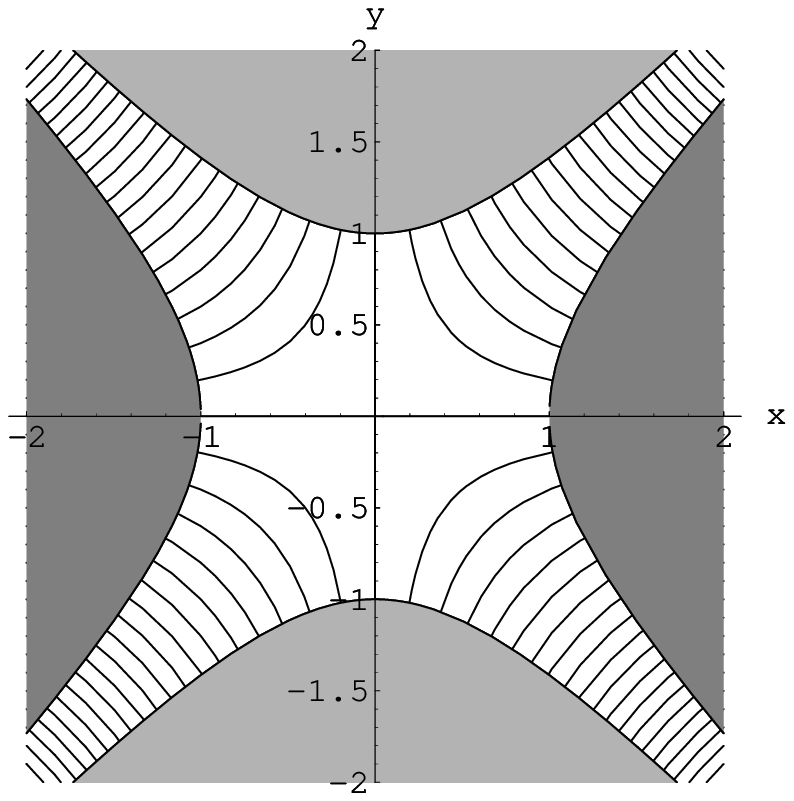,height=6cm}\hfill
\epsfig{file=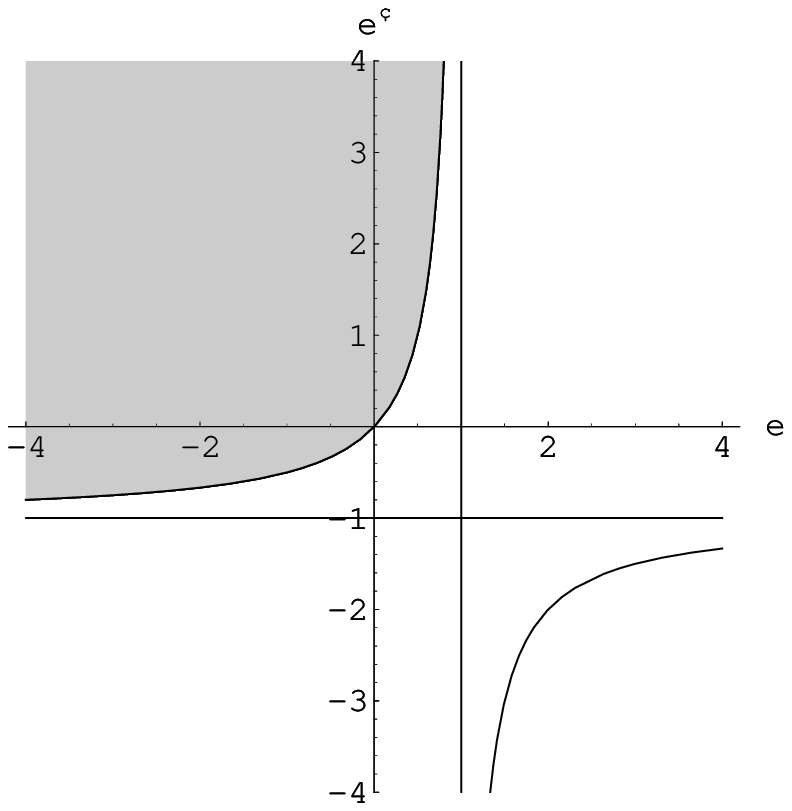,height=5.5cm}\hfill
\caption{Left: Field lines of the quadrupolar electrostatic potential
$V=e(x^2-y^2)/2$. Right: phase diagram for elastic dipoles in the 
$(e_0,e_1)$ plane. The stable region is the domain in dark. Macroscopic
strings get produced on the separatrix.}}

Let us start by considering a single direction $x$. The stability of
the potential $V$ depends crucially on the sign of the determinant of $Q$,
\be
D= \det(Q)=  e_1-e_0 -e_0 e_1
\ee
If (i) $D>0$ and $e_1-e_0 > -2$, the two eigenvalues are positive and the
origin is a global minimum. If (ii) $D>0$ and $e_1-e_0 < -2$ the two eigenvalues
are negative and the origin is a global maximum. Finally, in the
intermediate region (iii) where $D<0$, the two eigenvalues have opposite
sign and the origin is a saddle point. This phase diagram is shown
on Figure \ref{phasediag}. Note that in the limit of zero string tension, 
the only stable region would have been $e_1>0, e_0<0$, corresponding
to a stable electrostatic well for the two ends of the string. Thanks to
the elastic binding energy, 
the stability domain extends partly into the two quarters
$e_0 e_1>0$. If we now add a second direction $y$, due to the
tracelessness constraint on $h^i$, we are forced to choose opposite values
$(-e_0,-e_1)$ for the gradient along $y$. It is evident from Figure
\ref{phasediag} that there is no way to have the motion stable along the
two directions at the same time. More generally, due to the convexity
of the domain (i), there is no way to choose gradients $(e_0^i,e_1^i)$
along $n$ directions in the stable domain (i) while still maintaining
the tracelessness constraint $\sum_{i=1,n}(e_0^i,e_1^i)=(0,0)$. This
is consistent with the fact that the ground state of the string
has a {\it positive} polarizability. One may
ask whether choosing non-commuting electric gradients $(h_0,h_1)$
at the two ends may somehow stabilize the system, however investigation
of the crossed configuration $\Phi_0=h_0 (x^2-y^2)/2, \Phi_1=h_1 x y$ 
shows that this does not seem to be the case (see section \ref{nocom} in the 
case of the open string). We will see
in section \ref{stabmag} that stabilization can however be achieved
by using a magnetic field, or by modulating the strength of the
electric field.

In spite of the fact that 
this electrostatic configuration is globally unstable as it
stands, there is still a rather interesting phenomenon that takes place
at the critical line $D=0$: there the potential admits a degenerate valley,
corresponding to stretched strings of arbitrary size $L$ in the direction $x$,
\be
\label{macs}
X=L (e_0 \sigma -\pi)
\ ,\quad \mbox{for} \quad e_1 - e_0 - e_0 e_1 =0 \\
\ee
The physical origin of these states is clear: while the electrostatic
potential decreases quadratically (when $e_0 e_1>0$) 
like the  square of the length, the tensive energy of the 
non-relativistic string grows like the square length as well, and 
there exists a critical value of the electric {\it gradient} for which
the two forces cancel for any size. When this value is reached,
the strings become macroscopic and extend to infinity, thereby discharging
the condensator that created the electric field in the first place.

This phenomenon is very reminiscent of the critical electric {\it field
strength}
that appears in Born-Infeld electrodynamics, or equivalently 
for charged open strings in an electric 
field \cite{Fradkin:1985qd,Burgess:1986dw}: 
the electric potential that pulls on the
string ends  grows with the length 
of the string $L$ as $E L$, while the tensive energy
grows as $L/\alpha'$. At the critical value  $E_c=1/\alpha'$,
the two forces cancel and lead to strings of macroscopic
size in the direction of the electric field, and small effective tension. 
If one attempts
to go beyond $E_c$, the vacuum becomes unstable, tiny quantum strings 
get stretched to infinite distance very rapidly and discharge
the condensator, thereby maintaining $E\leq E_c$.
Such a regime has
been used to construct the NCOS theory, a theory of non-commutative
open strings decoupled from closed strings \cite{min}. It would
be very interesting to see if such a decoupling limit could be 
taken here as well, leading possibly to a theory of non-relativistic
open strings only (see \cite{Gomis:2000bd} 
for other attempts to define non-relativistic
limits of closed strings). Another interesting question is whether and
how these zero energy states can condense and backreact on the
background \eqref{wav}. This may be especially tractable
due to recent progress in open string field theory.

\subsection{Elastic dipole in a Penning trap}
\label{stabmag}
While the purely electrostatic configuration of the previous section
exhibited an interesting phase transition, it still suffers of a
global instability, that may shed doubt on the existence of these
macroscopic strings. We now switch on a magnetic field, and
investigate whether the same mechanism that allowed to trap charged
particles still succeeds in  stabilizing the string. 
We therefore consider our dipole model in the electrostatic and
magnetic fields
\be
\Phi_a = \frac12 h_a \left( x^2 + y^2 - \frac{2}{1+b^2} z^2 \right)
\ ,\quad B=b ~dx\wedge dy
\ee
Here, we have taken into account the modified harmonicity condition
\eqref{harm}, which arises in dimension 3 and higher. According
to our discussion in Section \ref{generalsetting}, this is the sole effect 
of $\alpha'$ corrections (we have set $2\pi\alpha' =1$).
Along the third direction $z$, the
dynamics is exactly the same as in the purely electrostatic case
up to a redefinition of $(e_0,e_1)$, hence stability in this direction
requires
\bse
\label{trnob}
\begin{gather}
\left(1+b^2 +2e_0\right)\left(1+b^2-2e_1\right) -
\left(1+b^2\right)^2 > 0 \\
\left(1+b^2\right)+e_0-e_1 >0 \label{trnobupperleft}
\end{gather}
\ese
In the transverse directions, the magnetic field leads to a modified
dispersion relation,
\be
\begin{pmatrix} 1 - e_0 - \om^2 & 2 i b \omega & -1 & 0 \\
-  2 i b \omega &  1 - e_0 - \om^2 & 0 & -1 \\
-1 & 0 & 1 + e_1 - \om^2 & 2 i b \omega  \\
 0 & -1 & -  2 i b \omega &   1 + e_1 - \om^2
\end{pmatrix}
\cdot
\begin{pmatrix}
x_L \\ y_L \\ x_R \\ y_R 
\end{pmatrix} (\om) = 0
\ee
At zero frequency, the effect of the magnetic field is nil, hence we
still have production of macroscopic string states at the critical
value of the electric gradients,
\be
\label{r0}
D= e_1-e_0 - e_0 e_1 = 0
\ee
However, crossing this line does not lead to an instability anymore,
as we now show. The discriminant of the dispersion relation (i.e.
the resultant $R$ of the polynomials $\det(Q)$ and $d\det(Q)/d\omega^2$)
factorizes into
\be
\label{r1}
\begin{split}
R=b^4 \left[ 4+ (e_0+e_1)^2 \right]^2 &
\left[ 4+ \left(e_0+e_1\right)^2+8 b^2(e_0-e_1-2) \right]^2 \\
&\qquad \left[\left((1+ b^2) - e_0 \right) 
\left((1+ b^2) + e_1 \right) -1 \right]
\end{split}
\ee
When any of these factors vanishes, two roots $\omega^2$
of $\det(Q)$ collide, and
may, or may not, leave the real axis. Experiment shows that 
the third factor gives rise to an accidental degeneracy but no
instability, while the fourth does indeed lead to the disappearance
of two positive real roots into the complex plane, and hence to a 
genuine instability. The crossing of \eqref{r0} only leads to the
appearance of one zero eigenvalue which becomes positive immediatly again.
If we restrict to the line $e_0+e_1=0$ for simplicity, 
we therefore have the following transitions:
\begin{enumerate}
\item $e_0<0$: stable motion;
\item $e_0=0$: one zero mode, macroscopic strings are created;
\item $0<e_0< b^2$: only real roots, stable motion;
\item $e_0= b^2$: two eigenvalues become imaginary;
\item $ b^2<e_0<2$: partially unstable motion;
\item $e_0=2$: one zero mode, macroscopic strings are created;
\item $2<e_0<2+ b^2$:  partially unstable motion;
\item $e_0=2+ b^2$:  two more eigenvalues become imaginary;
\item $e_0>2+ b^2$: completely unstable motion;
\end{enumerate}
\FIGURE{\label{phasediagmag}
\hfill\epsfig{file=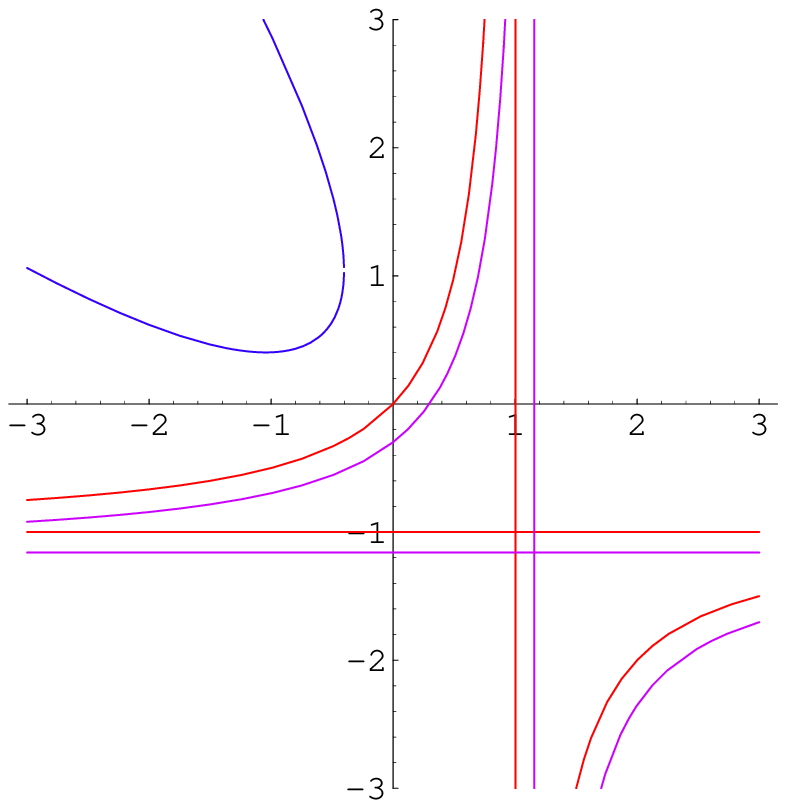,height=4.5cm}
\hfill\epsfig{file=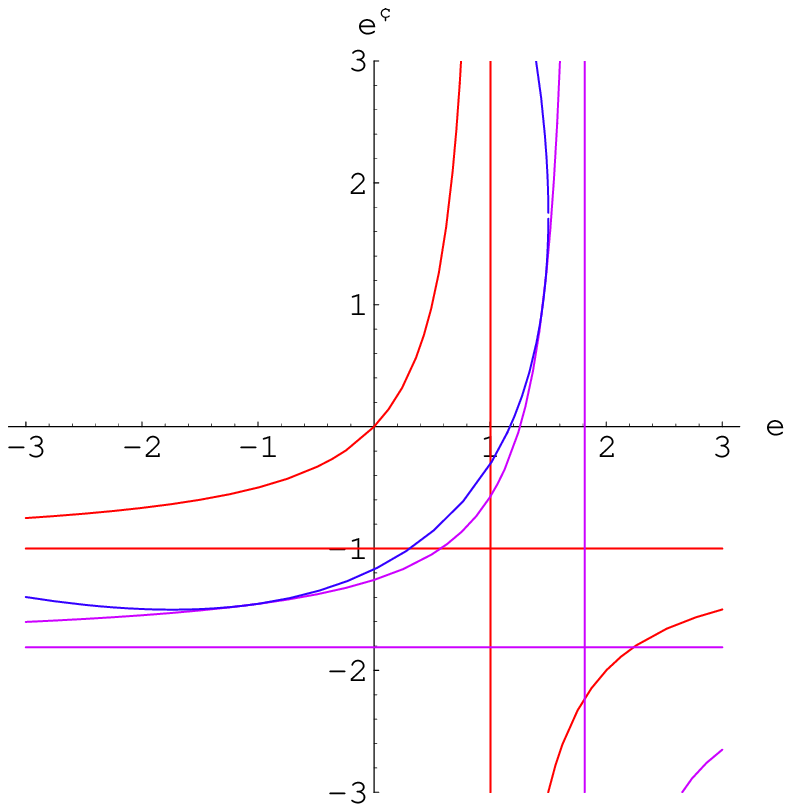,height=4.5cm}
\hfill\epsfig{file=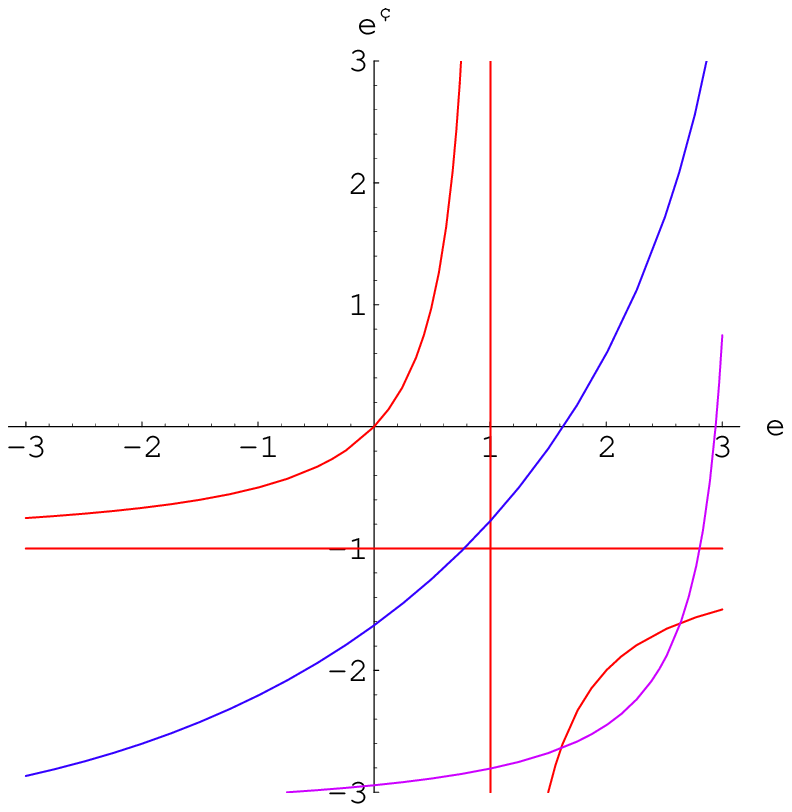,height=4.5cm}
\hfill
\caption{Phase diagram in the $(e_0,e_1)$ plane
for strings in a quadrupolar potential 
with small (left), intermediate (center) or large
(right) magnetic field; in purple, boudary of stability
domain; in red, line of production of macroscopic strings;
in blue, line of accidental degeneracy.}
}
Here we have assumed that $b^2<2$. If not, the stable 
region includes up to $e_0=2$, hence displays creation of
macroscopic strings twice. In addition, there is an additional
degeneracy at $e_0=1-1/(4 b^2)$, which can occur
either in the stable or unstable region. 
Examples of this phase diagram for several values of $b$ are given in
Figure~\ref{phasediagmag}. To conclude, in the presence
of a magnetic field, the stability region is enlarged to
the upper left quadrant of the hyperbola:
\bse
\label{condmag}
\begin{gather}
\left(1+ b^2 - e_0 \right) \left( 1+ b^2 + e_1 \right) - 1  > 0, \\
e_0 < 1+ b^2, \label{condmagupperleft}
\end{gather}
\ese                       
The stable region therefore includes all of the first critical line 
for creation of macroscopic strings, and extends into the formerly
unstable region up to 
\be D \sim  \frac{2-2e_0+e_0^2}{e_0-1}~ b^2  \qquad \mbox{for} \ \ b \ll 1
\label{criticalb} 
\ee
where the kinematic instability takes over. If $b^2>2$, part
of the second critical line for creation of macroscopic strings
is included into the kinematically stable region as well.
It is now possible to combine the two conditions \eqref{trnob}
and \eqref{condmag} to find a region with global stability
in all $(x,y,z)$ directions.


\subsection{Elastic dipole in a Paul trap}
\label{Paul}
As we recalled in Sec \ref{iontraps}, another standard way to trap 
ions in a quadrupolar electric field is to modulate the
electric field periodically in time at a resonant frequency. Our aim
in this section is to see to what extent the same mechanism can be
used to trap charged elastic dipoles. Let us start by reviewing
this mechanism in the point particle case. We thus consider 
a charged particle in a time-varying two-dimensional electric potential, 
\be
\Phi(x^+)= \frac12 \left[ h + f H\left(\frac{2\pi}{T}x^+\right)
\right] ( x^2 -y^2)
\ee
where $H(t)$ is a periodic function of $t$ with period $2\pi$.
After rescaling time, the equation of motion can be rewritten as
\be 
\label{mathieu}
 \ddot{X} + [\om^2 + \alpha^2 H(t)]~ X = 0
\ee
where $(\om^2,\alpha^2)=\pm (T/2\pi)^2 (h,f)$ for $X=x,y$
respectively. For $H(t)=\cos t$, this is known as the Mathieu
equation, whose solutions are well studied transcendental functions. 
The stability diagram of \eqref{mathieu} however depends little on the
details of the function $H(t)$, and can be studied much more easily
in the case of a rectangular signal \cite{vdpol},
\be
\label{rectsignal}
H(t)=1 ~\mbox{for}~ t\in[2n\pi,(2n+1)\pi[,\ ,\quad
H(t)=-1 ~\mbox{for}~ t\in[(2n+1)\pi,(2n+2)\pi[ \ .
\ee
Since $H(t)$ is periodic, the translation group must act by an
element $\rho$ of $Sl(2,\Real)$ on a fundamental basis of solutions
of the second order differential equation \eqref{mathieu},
\be
\begin{pmatrix} X_1 \\ X_2 \end{pmatrix} (t+2\pi)
= \begin{pmatrix} a & b \\ c& d \end{pmatrix}
\begin{pmatrix} X_1 \\ X_2 \end{pmatrix} (t)\ ,\quad
ad-bc=1\ ,\quad W(X_1,X_2)=0
\ee
where $W(X_1,X_2)=X_1 \p_t X_2 - X_2 \p_t X_1$ is the Wronskian,
conserved in time. The eigenvalues of this $Sl(2,\Real)$ element 
can be written as $e^{\pm 2\pi i \mu}$ where the complex-valued
Floquet exponent $\mu$ is computed from $\cos 2\pi \mu= (a+d)/2$. The motion is
stable iff $\mu$ is real, i.e. $|\cos(2\pi \mu)| \leq 1$. 
For the rectangular signal \eqref{rectsignal}, 
the matrix $\rho$ can be computed
straightforwardly by matching $X$ and its derivative at $t=0$ and $t=\pi$.
The Floquet exponent turns out to be determined by \cite{vdpol}
\be
\cos(2\pi \mu) =
\cos x_1 \cos x_2 - \frac12 \left( \frac{x_1}{x_2} + \frac{x_2}{x_1}
\right)\sin x_1 \sin x_2 
\ee
where 
\be
x_1 = \pi \sqrt{\om^2+\alpha^2}\ ,\quad
x_2 = \pi \sqrt{\om^2-\alpha^2}\ ,\quad
\ee
The domain of stability $|\cos(2\pi\mu)| \leq 1$ in the $(\om^2,\alpha^2)$
plane is depicted in Figure \ref{stability}. For a small ripple $\alpha\gg
\om$ and $\om^2>0$, the motion is generically stable except when 
$\omega$ is integer or half-integer. At these values, the
forcing is in resonance with the proper oscillation modes and
an infinitesimal perturbation $\alpha$ is sufficient to 
destabilize the motion. This unstable region appears as a cusp
on Figure \ref{stability}, with boundaries well approximated at small
$\alpha$ by
\bea
\om^2 &=& n^2 
+ \left\{ \begin{array}{c}
  -\frac{1}{4\,n^2}\alpha^4 +
  \frac{  \left( -9 + n^2\,{\pi }^2 \right) }{48\,n^6} {\alpha}^8
\\
  \frac{3}{4\,n^2}{\alpha}^4  - 
  \frac{     \left( 21 + n^2\,{\pi }^2 \right) }{48\,n^6} {\alpha}^8 
\end{array} \right\} + {\cal O}(\alpha^{12})\ ,
\qquad n\in\mathbb{N}^*\\ 
\om^2 &=& 
\left(n+ \frac12\right)^2 
+ \left\{ \begin{array}{c}
  \frac{2}{\pi(2n+1)} \alpha^2
+ \frac{\pi^2 (2n+1)^2-12}{(2n+1)^4 \pi^2} \alpha^4 \\
- \frac{2}{\pi(2n+1)} \alpha^2
+ \frac{\pi^2 (2n+1)^2-12}{(2n+1)^4 \pi^2} \alpha^4
\end{array} \right\} +  {\cal O}(\alpha^{6})\ ,\quad n\in \mathbb{N}
\eea
Conversely, an unstable motion at $\alpha=0$ can be made stable by
switching on a small perturbation. Indeed, there appears to be
a stable region extending in the $\om^2<0$ domain above the line
attaching at 0,  well approximated by
\be
\om^2 = -\frac{\pi^2}{12} \alpha^4 + \frac{\pi^6}{1512} \alpha^8
-\frac{107 \pi^{10}}{9979200}\alpha^{12} + {\cal O}(\alpha^{16})
\ee
It is therefore possible to choose the period and amplitude
of the forcing $(f,T)$ so that both
$(\omega^2,\alpha^2)$ and $(-\omega^2,-\alpha^2)$ are in the
stable region, hence the motion is stable in both directions
$x$ and $y$.

\FIGURE{\label{stability}
\hfill\epsfig{file=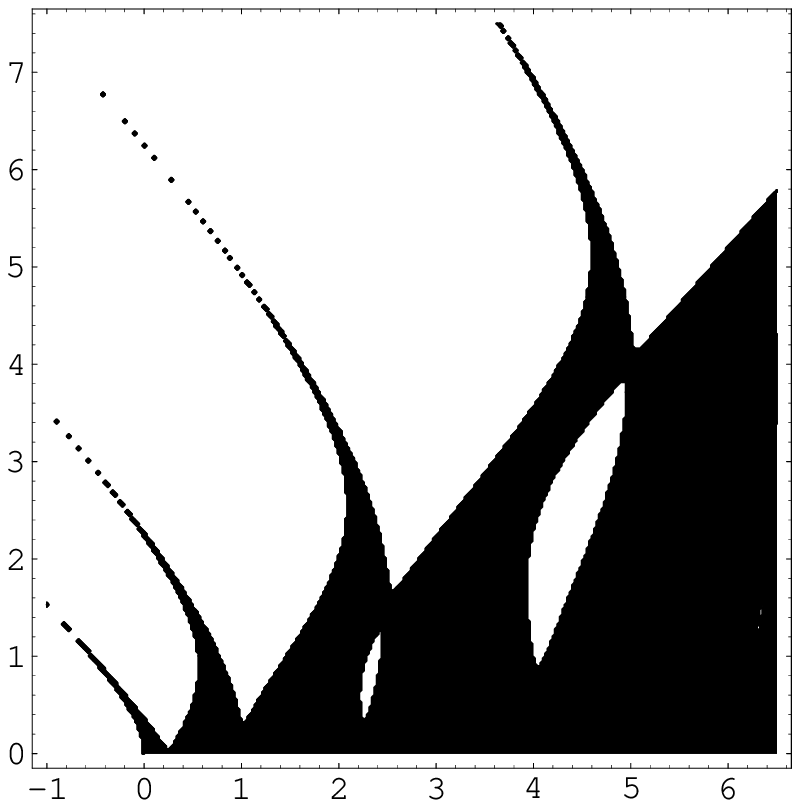,height=4.5cm}
\hfill\epsfig{file=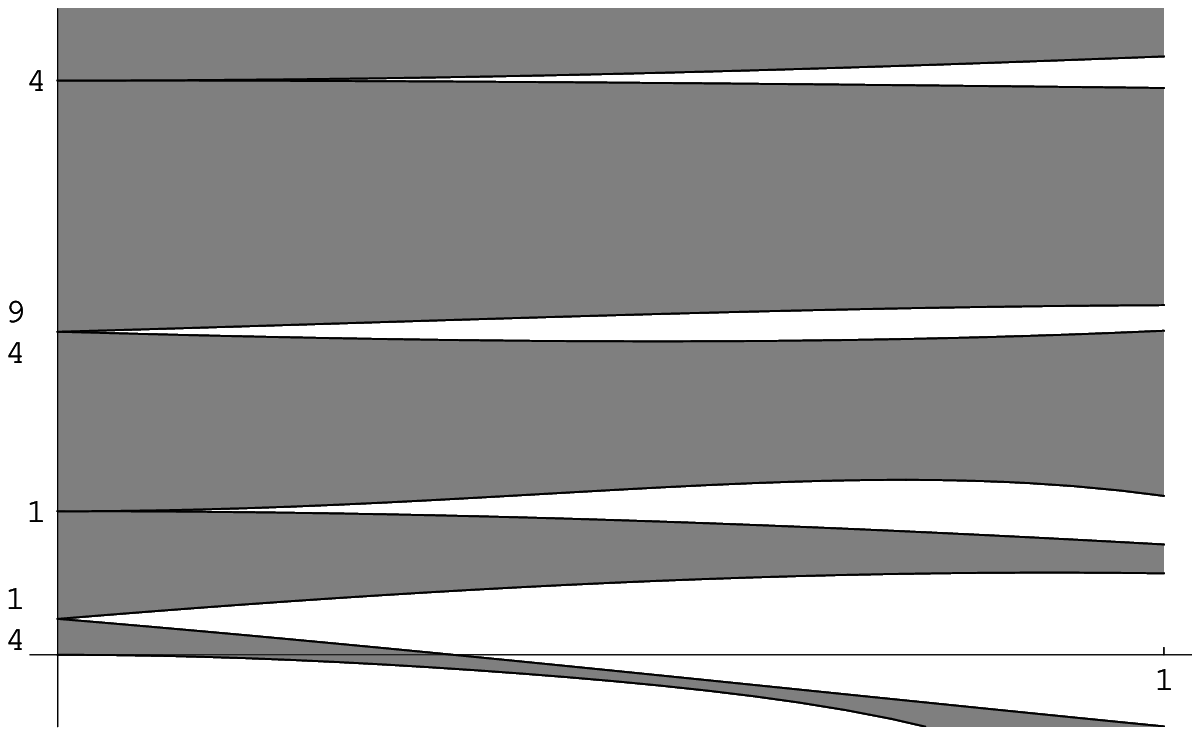,height=4.7cm}
\hfill
\caption{Left: stability diagram for an harmonic oscillator with
frequency modulated by a rectangular signal,  
in the $(\omega^2,\alpha^2)$ plane. The stable region appears shaded, 
notice that it extends into the usually unstable regime $\om^2<0$. 
Right: zoom on the region  where $\alpha^2\ll 1$, axes are exchanged.
Instabilities set in at half-integer $\omega$, even with an infinitesimal
amplitude of modulation.}
}

Let us now consider the case of an elastic dipole in a modulated
quadrupole field. For simplicity, we assume that the 
same modulating frequency $T_0=T_1=T$ is applied on either end of the open
strings, with opposite amplitudes $f_0=-f_1=f$. The equation of motion
\be
\begin{pmatrix} \ddot{x}_L \\ \ddot{x}_R \end{pmatrix}
+ \frac{1}{\pi}
\left[ \begin{pmatrix}
1- \pi p^+h_0 & -1 \\
-1 & 1+ \pi p^+ h_1
\end{pmatrix}
- p^+ f H\left(\frac{2\pi}{T} x^+\right)
\begin{pmatrix}
1 & \\ & 1
\end{pmatrix}
\right]
\begin{pmatrix} x_L \\ x_R \end{pmatrix} = 0
\ee
can then be reduced to \eqref{mathieu} for each of the proper directions of 
the constant matrix $Q$, with
\bea
\om^2_{\pm} &=& \left(\frac{T}{2\pi p^+}\right)^2  \frac{1}{\pi}
\left( 1+ \frac{e_1-e_0}{2} \pm \sqrt{1+\frac{(e_0+e_1)^2}{4}} \right) \\
\alpha^2 &=& \left(\frac{T}{2\pi p^+}\right)^2 f
\eea
where the two signs correspond to the two modes of $Q$ in the $x$
direction. In the limit of small tension, the two eigenmodes
$\om^2_{\pm}$ become
\be
\omega_+= \left(\frac{T}{2\pi p^+}\right)^2 \frac{1}{\pi}
\left( -e_0 + 1  + {\cal O}(1/h) \right) \ ,\quad
\omega_-= \left(\frac{T}{2\pi p^+}\right)^2 \frac{1}{\pi}
\left( e_1 + 1 + {\cal O}(1/h) \right)
\ee
The eigenmodes in the $x$ and $y$ directions are therefore not
opposite anymore but shifted together upward. Nevertheless it is clear
that for small enough tension one will still be able to choose
the perturbation $f$ such that both directions are stabilized.

\subsection{Elastic dipole in an adiabatically varying electric field}
After having discussed the motion of an elastic dipole in a resonant 
quadrupolar case, we now consider a dipole in a time-dependent
quadrupolar potential $\Phi_a(x^+)=h_a(x^+)(x^2-y^2)/2$, which we
assume to vary adiabatically between two non-zero
asymptotic values $h_a(\pm\infty)$. 
In the particular case $h_0(x^+)=-h_1(x^+)=h(x^+)$, the mass matrix
$Q(x^+)$ can be diagonalized independently of time, yielding two
decoupled harmonic oscillators with time-dependent frequency,
\be
  \ddot{x}_\pm + \omega^2_\pm(\tau) x_\pm = 0
\ee
with $\omega^2_\pm = p^+ h(x^+) \pm 1$, $x_\pm=x_L \mp x_R$. This problem
can be solved for an arbitrary profile $h(x^+)$ by separating the modulus
and phase of, say, $x_\pm=s_\pm e^{i \gamma_\pm}$. 
The imaginary part of the equation
relates the rate of phase variation to the modulus through $s_\pm^2 
\dot\gamma_\pm= \text{cte}$ which can be set to 1. 
The real part gives a non-linear differential equation for the 
modulus,
\be
\label{lewis}
 \ddot{s}_\pm + \omega^2_\pm(\tau) s_\pm =
 \frac{1}{s_\pm^3}
\ee
The complete quantum mechanical S-matrix can then be found from the
knowledge of $s_\pm$ at late times, where \eqref{lewis} reduces to the equation of
motion deduced from the
Hamiltonian of conformal quantum mechanics \cite{lr}. This result
does not rely on any adiabaticity assumption, although determining
$s_\pm(+\infty)$ may not be feasible analytically.

In the general case however, the proper directions of $Q$ change with
time, and it no longer helps to diagonalize the Hamiltonian. Instead,
one may eliminate $x_L$ (say) and obtain a $4^\text{th}$-order differential equation
for $x=x_R$,
\be
\label{fourthorder}
\pi x^{(4)} + (\tr Q) \ddot{x}
+ \det Q =0
\ee
where we recall that  $\tr Q = 2 + e_1 - e_0$ and 
$\det Q = e_1 - e_0 - e_1 e_0$. It is not possible anymore to separate 
the phase and modulus of $x=s e^{i \gamma}$ in general 
(although the imaginary part of \eqref{fourthorder} can be integrated once
after multiplying by the integrating factor $s(\tau)$), 
however the problem can be
solved in the adiabatic regime $\dot h_a \ll h^{3/2}$. Following the
standard WKB method, we rescale $t\to t/\epsilon$ and 
$\gamma\to\gamma/\epsilon$. At leading order, \eqref{fourthorder} identifies
the rate of phase variation
with one of the instantaneous proper frequencies,
\be
\label{adiaborderzero} 
\dot{\gamma} = \omega (\tau) = \pm \sqrt{\frac{\tr Q}{2\pi} 
\pm \frac1{\pi}\sqrt{1 +\frac{1}{4} (e_0+e_1)^2 }}\ , 
\ee
hence $\gamma(\tau)=\int_0^\tau \om(\tau') d\tau'$, 
while the amplitude  is given at next-to-leading order by 
\be s(\tau) =  \frac{s(0)}{\left({\cal A}(\tau)\right)^{1/2}} \exp \int_0^\tau 
\frac{ip^+(\dot{e}_1(\tau') - \dot{e}_0(\tau'))\om(\tau')}{\pi {\cal A}(\tau')} d\tau'
\ee
where
\be {\cal A}(\tau) = 2\left(\frac{i}{\pi}(e_1(\tau) -e_0(\tau) + 2) \om(\tau)
- 2i\om(\tau)^3 \right) \ee
and $\omega$ denotes one of the branches of \eqref{adiaborderzero}.
The approximation remains valid as long as all eigenvalues stay separate
from each other. In particular it breaks down at the line of production
of macroscopic dipoles where two eigenvalues collide and leave the
real axis. It would also break down if the electric gradient $h_a$ was 
switched off at $\pm\infty$, except in the 
proper direction $x^+$ which remains confined due to the string tension.

\section{Open strings in a null quadrupolar electric field}
\label{nullquad}
We now proceed with the first quantization of a string in a 
quadrupolar electrostatic potential $\Phi_a={h_a}_{ij} x^i x^j/2$,
where ${h_a}_{ij}$ are constant traceless matrices, independent of $x^+$. 
For now, we do not assume any particular commutation relation
between $h_0$ and $h_1$.

\subsection{First quantization in a constant quadratic potential}

When decomposing $X$ in left and right movers \eqref{lrmovers},
the boundary condition \eqref{bcquad} translates into the system
of linear equations 
\be 
\label{diffdiff}\begin{aligned}
\dot{f}(\tau) - \dot{g}(\tau) &+ p^+ h_0 \bigl(f(\tau)+g(\tau)\bigr) 
+ B \bigl(\dot{f}(\tau)+\dot{g}(\tau)\bigr) = 0 \\
T^2\dot{f}(\tau) - \dot{g}(\tau) &+ p^+ h_1 \bigl(T^2f(\tau) + g(\tau)\bigr)
+ B \bigl(T^2\dot{f}(\tau)+\dot{g}(\tau)\bigr) = 0 
\end{aligned}
\ee
where $T^2$ denotes the translation operator\footnote{Its square root 
$Tf(\tau)=f(\tau+\pi)$ will be of use later on.}, 
$T^2f(\tau)=f(\tau+2\pi)$.
In order to avoid cluttering, we have absorbed a factor $p^+$ 
into the gradients $h_a$, and omitted the $d\times d$ indices. Such non-local
differential problems have been studied in the mathematical literature 
under the name of linear differential difference equation of neutral 
type \cite{hale}. They are in contrast to the case of closed strings
in gravitational waves, where the classical dynamics reduces to 
an ordinary differential equation for each mode of the Fourier
expansion of the periodic coordinates $X(\sigma)$. They can nevertheless
be solved rather straightforwardly using Fourier (or Laplace) analysis.

\subsubsection{Neutral string}
\label{neutral}
Before proceeding to the general case, let us first discuss the
neutral case of an open string starting and ending 
on the same brane $h_0=h_1:=h$.
Taking the difference of the two equations above,
we get
\be
(T^2-1)(\dot{f}+ p^+ hf+B\dot{f})=0
\ee
Hence $\dot{f}(\tau)+p^+ hf(\tau)+B\dot{f}$ has to be a periodic function of $\tau$, with the
Fourier series expansion (we assume $[e,B] = 0$),
\be \label{Fourierexp}
\dot{f}(\tau)+\frac{e}{\pi} f(\tau)+B\dot{f}(\tau) =
a_0 + \sum_{n\neq 0} \left((1+B)+\frac{i}{n \pi} e\right) {\cal M}_n a_n e^{-in \tau} =
\dot{g}(\tau)-\frac{e}{\pi}g(\tau)-B\dot{g}(\tau).
\ee
We chose a somewhat awkward normalisation of the $a_n$'s in order to get the correct commutation relations
at the end.
\eqref{Fourierexp} can be easily integrated to yield the normal mode expansion:
\be \label{modexpeq}
\begin{split}
X(\tau,\sigma) = &{\cal M}^u_+ 
 \frac{\pi(1-B)}{e}  e^{\frac{e}{\pi(1-B)}(\tau-\sigma)} a_+ - {\cal M}^u_-
\frac{\pi(1+B)}{e} e^{-\frac{e}{\pi(1+B)}(\tau+\sigma)} a_-  \\
& + \sum_{n\neq 0} {\cal M}_n  \frac{i}{n} \left[ e^{-in(\tau+\sigma)}+ 
\frac{in\pi(1+B) -e}{in\pi(1-B) +e} e^{-in(\tau-\sigma)}
\right] a_n \end{split}
\ee
where $d\times d$ matrices are still understood.
The normalisation coefficients are given in Appendix \ref{norm}.
The expansion therefore remains integer modded as in the $e=0$ case, with the
apparition of exponentially varying zero-modes $(a_+,a_-)$.
With the normalisation chosen, we have the standard commutation relations:
\be [a^i_n, a^j_m] = n \delta^{ij} \delta_{n,-m} 
\qquad [a^i_+, a^j_-] = i\frac{1}{\pi} e^{ij} \ee
The Hamiltonian \eqref{lchamil} can be easily computed:
\be
p^+ {\cal H}_{lc} =a_+ a_- + \sum_{n=1}^\infty a^\dag_n a_n - \frac{D}{24}
\ee

\subsubsection{Charged string}
Let us now come back to the general case with $e_0\neq e_1$, in
the absence of a magnetic field for simplicity. Decomposing
$f$ and $g$ into their Fourier modes, Equation \eqref{diffdiff} 
becomes a linear system
\be
\label{dispmatrix}
\begin{pmatrix} -i\omega + \frac{e_0}{\pi} & i\omega + \frac{e_0}{\pi} \\
(-i\omega + \frac{e_1}{\pi}) e^{-2i\pi \omega} & 
i\omega + \frac{e_1}{\pi} \end{pmatrix} 
\begin{pmatrix} \hat{f}(\omega) \\ \hat{g}(\omega) \end{pmatrix} = 0 
\ee
The compatibility of this linear system puts a quantization condition
on the energies $\omega$, which can be rewritten as the secular equation
\be
\label{fdisp}
\det \left[ \left((\pi\omega)^2+e_1 e_0\right) \sin(\pi \omega) 
- (e_1 - e_0) \pi\omega \cos(\pi \omega)  \right] =0
\ee

Now let us assume that $e_0$ and $e_1$ commute
(the case where they do not will be briefly considered
in section \ref{nocom}). 
The two matrices can then be diagonalized simultaneously and we can restrict ourselves
to a potential of the form $\frac12 (x^2 -y^2)$ where $x$ and $y$ are two spatial directions. 
The problem thus decouples into two one-dimensional problems
related by inverting $(e_0, e_1) \to (-e_0, -e_1)$.
The dispersion relation \eqref{fdisp} thus reads
\bse
\be
\label{secu}
e^{2 \pi i\om} = \frac{i\pi \om + e_0}{i\pi \om - e_0}
\frac{i\pi \om - e_1}{i\pi \om + e_1}  
\ee
or equivalently
\be
\tan(\pi \om) = \frac{ (e_1-e_0) \pi \om}{(\pi\om)^2 + e_0 e_1}
\ee
\ese
To find solutions of \eqref{secu} graphically, one has to look at the intersection
points of the ``Lorentzian curve" on the \rhs with
the $\tan$ curve on the \lhs. The secular equation 
selects one real eigenmode $\omega_n$ in any interval
$[n\pi,(n+1)\pi[$, $n\in \mathbb{N}^*$, with wave function
\be
\label{genmode}
X_n(\tau,\sigma)= {\cal N}_n \frac{i}{\omega_n} \left( 
e^{-i\om_n(\tau+\sigma)} + 
\frac{i\pi\om_n-e_0}{i\pi\om_n+e_0} e^{-i\om_n (\tau-\sigma)} \right) a_n 
+ h.c. \qquad n>0
\ee
where the normalization factor ${\cal N}_n$, given in Appendix \ref{norm},
ensures that the commutators are indeed $[a_n,a_m^\dag]= \om_n \delta_{n,m}$. 

\FIGURE{ \label{seg11}
\hfill\epsfig{file=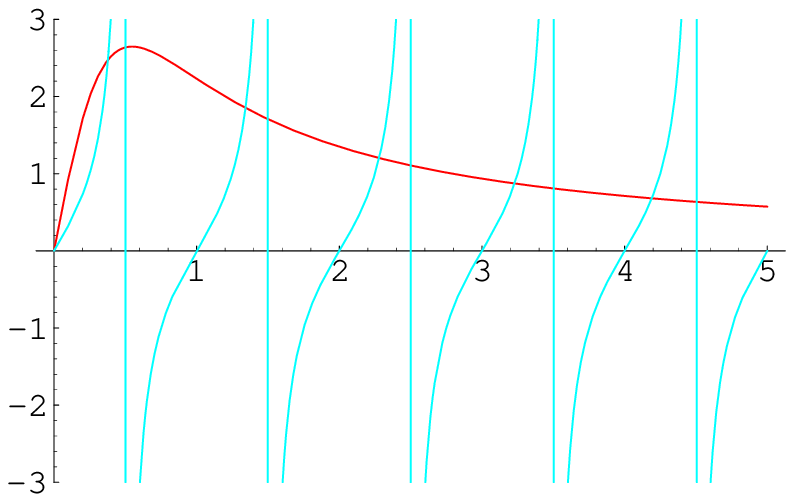,height=3.5cm, width=6cm}
\hfill\epsfig{file=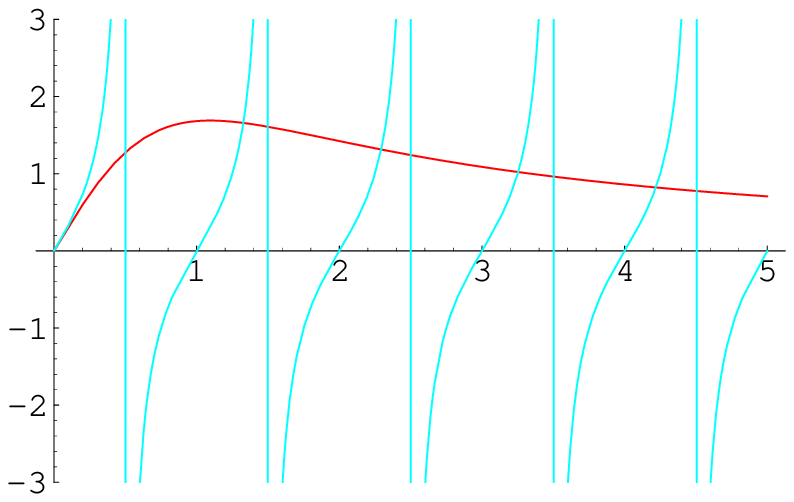,height=3.5cm, width=6cm}
\hfill
\caption{Graphical solution of the open string dispersion relation for
a one-dimensional harmonic potential. Left: $D>0$, $e_1-e_0>-2$, all roots
are real. Right: $D<0$, $e_1 - e_0 > -2$, two roots have collided at $\om=0$ and left
into the complex plane.}
}

\FIGURE{ \label{seg12}
\hfill\epsfig{file=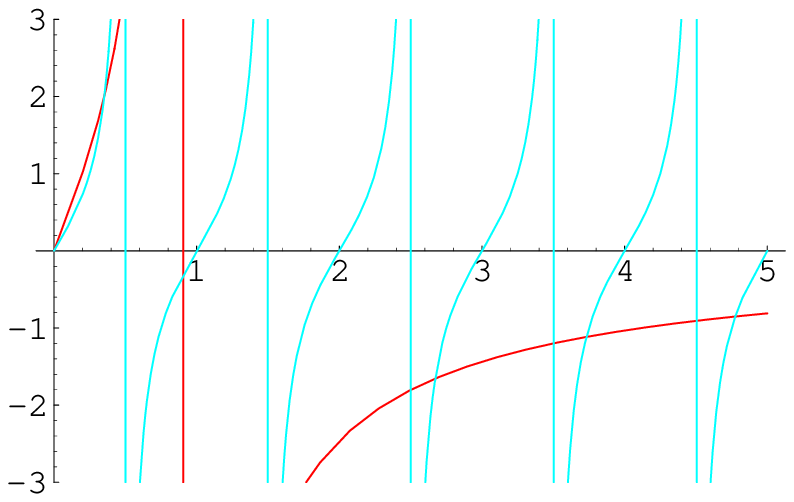,height=3.5cm, width=6cm}
\hfill\epsfig{file=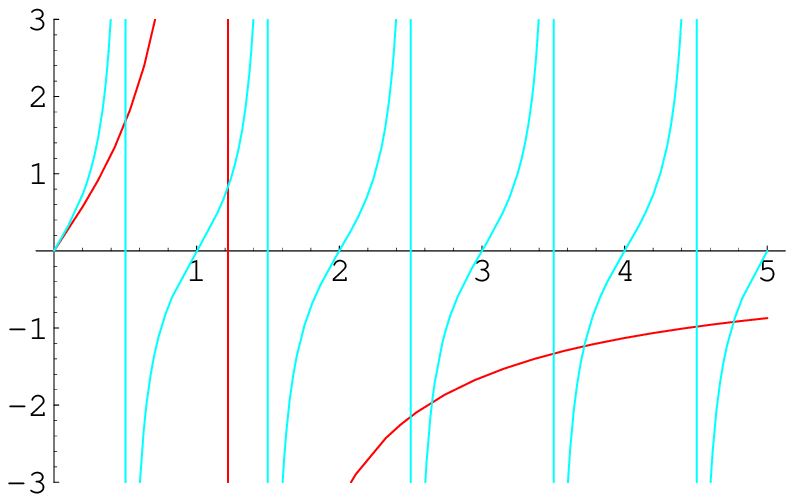,height=3.5cm, width=6cm}
\hfill
\caption{Graphical solution of the open string dispersion relation for
a one-dimensional harmonic potential. Left: $D<0$, $e_1-e_0<-2$, the root in the interval
$[0, 1]$ corresponds to the first excited level.
Right: $D>0$, $e_1-e_0 < -2$, two more roots have collided at $\om=0$ and left
into the complex plane.}
}

The situation for the first segment $[0,1[$ is more subtle however (see Figures \ref{seg11} 
and \ref{seg12}):
\begin{itemize}
\item $D=e_1-e_0 - e_0 e_1 > 0$ and $e_1 - e_0 > -2$

\noindent There is a single real root
$\omega_0 \neq 0$ in $]0,1[$. As shown in Appendix \ref{proof}, 
all the roots are then real
and the motion is stable.
The wave function $X_0$ is the same as \eqref{genmode} with $n=0$.

\item $D=0$ and $e_1 - e_0 > -2$

\noindent As $D$ approaches 0, the real root $\omega_0$ collides with its opposite
$-\omega_0$ ,
leading to a pair of canonically conjugate modes with
vanishing frequency. Their wave function can be obtained from 
the $\omega_n\to 0$ limit of \eqref{genmode}, and reads
\be
X_c(\tau,\sigma)={\cal N}_0^c (1-\frac{e_0}{\pi} \sigma)(x_0+p_0 \tau)
\ee
where again the normalization (see Appendix \ref{norm}) gives the canonical
brackets, $[x_0,p_0]~=~i$.
Those are exactly the macroscopic strings described in the previous
section, occurring at a critical value of the gradient of the electric field.
For small deviation away from the critical gradient, their energy is
given by
\be
\label{om0}
\omega_0= \sqrt{\frac{3(1-e_0)}{\pi^2(2-3e_0+e_0^2)} D}
\ee

\item $D < 0$ (note that the line $e_1 - e_0 = -2$ is contained in this domain)

\noindent As $D$ becomes negative, these two roots leave the real axis and
become purely imaginary, $\omega_0^\pm = \pm i k_0$.
This corresponds to unstable modes, with wave function
\be
\label{genmode0}
X_\pm(\tau,\sigma)= {\cal N}^u \frac{1}{k_0} \left(
e^{\pm k_0(\tau+\sigma)} + 
\frac{\mp\pi k_0-e_0}{\mp \pi k_0+e_0} e^{\pm k_0 (\tau-\sigma)} \right) a_{\pm}
\ , \quad [a_+,a_-]= ik_0
\ee

\item $D > 0 $ and $e_1 -e_0 < -2$

\noindent When crossing the $D=0$ curve in the region $e_1 -e_0 < -2$,
the two roots $\pm \omega_1$ collide, hence two new macroscopic modes
appear  and then become unstable. They behave like unstable modes described above in the case
where $D<0$.
\end{itemize}

Altogether, we have thus found exactly the same phase diagram as
in our dipole model, implying that the afore-mentioned instabilities
have indeed a low-energy origin.
Having found the normal modes, we can now carry out a standard
mode expansion, focusing on the region of the stability diagram where $e_1 - e_0 > -2$,
\bse \label{modexpansion}
\begin{gather}
X(\tau,\sigma) = 
\sum_{n=1}^{\infty} X_n
+ \begin{cases}
D>0:& X_0  \\
D=0:& X_c  \\
D<0:& X_+ + X_- 
\end{cases}
\end{gather}
\ese
with canonical commutation relations,
\be[a_n,a^\dag_m]~=~\om_n \delta_{n,m},\qquad [x_0,p_0]~=~i, \qquad [a_+, a_-]~=~ik_0.\ee
The contribution of the $X$ coordinate to the light-cone Hamiltonian 
is easily evaluated in terms of these variables, and reads
\be
\label{hamil}
p^+ {\cal H}_X= \sum_{n=1}^{\infty} a^\dag_n a_n
+\left\{ \begin{matrix}
D>0:& a^\dag_0 a_0 \\
D=0:& p_0^2 \\
D<0:& a_+ a_-
\end{matrix}\right\}
+ E_{X}
\ee
The vacuum energy $E_X$ has been evaluated using contour integral methods
in \cite{Arutyunov:2001nz}. The contribution
of the boson $Y$ can be obtained from \eqref{hamil} by simply reversing the
sign of $(e_0,e_1)$ (note that $D$ is not invariant under this operation).

\subsubsection{Non commuting electric gradients}
\label{nocom}

The dispersion relation \eqref{fdisp} cannot be solved as easily as before,
but we can examine the following crossed configuration:
\be
h_0 = h \begin{pmatrix} 1 & \\ & -1 \end{pmatrix}\ ,\quad
h_1 = h' \begin{pmatrix}  & 1 \\ 1 &  \end{pmatrix}
\ee 
The secular equation still factorizes into
\be\label{sec2}
\tan(\pi \om) = \pm \frac{\pi \om~ \sqrt{e^2 + e^{'2}}}{\sqrt{(\pi\om)^4 + e^2 e^{'2}}}
\ee
The structure of roots is similar to the commuting case, except that the
critical line is now at $D=e^2+e^{'2}- e^2 e'^2=0$,
at which the slope at $\omega=0$ of the \lhs of \eqref{sec2}
coincide the \rhs (with the $+$ sign).
As before, for all $n>0$ there exists a real root in $[n,(n+1)[$. 
For the branch with $-$ sign, the slope
at $\om = 0$ will always be negative and there will
never be an additional zero in the interval  $[0,1[$. However, there are 
always two complex roots $\hat\omega_0=\pm i\tilde k_0$
, that is to say, two unstable modes $\hat a_+$ and $\hat a_-$ (we
use a $\hat{}$ for oscillators of the minus branch, nothing for the
plus branch).
For the positive branch, the situation is comparable to the commutative case.
If $D>0$, there also exists a real root $\omega_0$ in $[0,1[$.
As $D\to 0$, this roots collides with its opposite and leaves the
real axis. For $D<0$, it becomes a pair of complex conjugate roots
$\omega_0=\pm ik_0$ corresponding to the instability described 
in the previous section. At $D=0$, we have macroscopic strings 
with arbitrary length,
\be
\begin{pmatrix} X_c\\Y_c \end{pmatrix}=
\sqrt{\frac{3(e^2- \pi)}{2 e^3 \pi^2}}  (x_0 + p_0 t)
\begin{pmatrix}
\frac{1-\frac{e}{\pi} \sigma}{\sqrt{e -1}} \\
\frac{1+\frac{e}{\pi} \sigma}{\sqrt{e +1}} 
\end{pmatrix}
\ee
and fixed angle $X_c/Y_c=\sqrt{(e+1)/(e-1)}$.
The Hamiltonian can be evaluated on this configuration and reads
\be
p^+ {\cal H}= \sum_{n=1}^{\infty} a^\dag_n a_n 
+ \sum_{n=1}^{\infty} \hat a^\dag_n \hat a_n
+\left\{\begin{matrix}
D>0:&  a^\dag_0 a_0 \\
D=0:& p_0^2 \\ 
D<0:& a_+ a_-
\end{matrix} \right\} 
+ \hat a_+ \hat a_-
+E_{vac}
\ee
Non-commuting electric gradients therefore lift part of the degeneracy
of the commuting case, but do not prevent the kinematic instability
to take place.

\subsection{Open string in a Penning trap}
\label{opstrpenning}

While the sign of the quadrupolar potential was chosen so that the
direction $z$ was confining, the transverse $(x,y)$ plane corresponds
to unstable directions of the electric field. As we discussed in
\ref{stabmag}, this can be stabilized 
by adding a constant magnetic field in the $(x,y)$ plane. 
For simplicity we assume that $e$ and $B$ commute.
The linear system \eqref{dispmatrix} becomes
\be \begin{pmatrix} -(1+B)i\pi\omega + e_0 & (1-B)i\pi\omega + e_0 \\
(-(1+B)i\pi\omega + e_1) e^{-2i\pi \omega} & (1-B)
i\pi\omega + e_1 \end{pmatrix} 
\begin{pmatrix} \hat{f}(\omega) \\ \hat{g}(\omega) \end{pmatrix} = 0 
\ee
The dispersion relation splits into two branches,
\be
\label{secu2}
\tan(\pi \omega) = \frac{ \pi\omega (e_1-e_0) }{(1+b^2)(\pi\omega)^2 \pm
b (e_0+e_1)(\pi\omega) + e_0 e_1}
\ee
where the two signs correspond to the left and right-handed
circular polarizations. Reversing $\omega$ exchanges the two
branches, so that the two polarizations are canonically conjugate;
we can therefore focus on the upper sign only.  As before,
a phase transition occurs when the tangents of the two
curves on either side of \eqref{secu2} become equals. As $b$ couples only
at  higher order in $\omega$, the transition line remains
unaffected at $D=e_1-e_0- e_0 e_1=0$. In contrast to \eqref{secu} however,
only a single root collides with $\omega=0$, as can be seen by expanding
the dispersion relation at $\omega=0$ close to the transition line
$D\to 0$, 
\be
\label{Dzeroexpand}
\frac{(e_0-1)D}{e_0^2} (\pi\om)
+\frac{(e_0-2)b}{e_0} (\pi\omega)^2 
+\frac{(1-3b^2)(3-3e_0+e_0^2)}{3e_0^2} (\pi\om)^3 + {\cal O}(\om^4)= 0
\ee
When $D\to 0$, one of the roots collides with $0$, but simply crosses it
as $D$ changes sign. There is therefore no instability associated 
with the phase transition at $D=0$, instead a creation of
static macroscopic strings takes place as in the absence of 
a magnetic field. This is not surprising, since the effect of the magnetic
field should only be felt in dynamical situations. On the other hand,
an instability does take place as one goes further away from $D=0$:
indeed, by looking at the next term in the expansion \eqref{Dzeroexpand},
one sees that the two roots $\omega_0$ and $\omega_1$ collide and
leave the real axis at a critical value
\be
b_{c}^2 = \frac{4 (e_0-1)(3-3e_0+e_0^2)}{3 e_0^2(e_0-2)^2} D \ ,
\quad D\ll 1
\ee
This is qualitatively the same as in our dipole model \eqref{criticalb}.
As in the case of the dipole model in section \ref{stabmag}, the stability region
is enlarged so as to include a part of the macroscopic string creation line and to allow
global stability region as regards $x$ and $y$ coordinates. 
Unfortunately, the precise critical line at finite $D$ does
not seem to be computable explicitely. At any rate, this demonstrates
that the same set-up that allowed to stabilize elastic dipoles 
in a 3D quadrupolar potential also works for the full open string.

\subsection{One-loop amplitude and open-closed duality}
\label{closedstr}
Just as gravitational plane waves exhibit no particle production
nor vacuum polarization, the same is true for electromagnetic plane waves.
A simple way to see this is to return to the Coleman formula for
the one-loop vacuum free energy, written in light-cone variables,
\be
\label{coleman}
A = \frac12 V_{+-}
\sum_i \int_0^{\infty} \frac{dt}{t} 
\int \frac{dp^+ dp^-}{(2\pi)^2} 
\exp\left( i (p^+ p^- + m_i^2) t \right)
\ee
where we included the transverse momenta and string excitation levels
in the label $i$, and denoted by $V_{+-}$
the volume of light-cone directions.
The integral over $p^-$ enforces a delta function
$\delta(t p^+)$, which implies that only $p^+=0$ states contribute
to the vacuum energy: but these states are insensitive to the presence
of the null electromagnetic field, hence the vacuum energy reduces to
that of flat space. Equivalently, in order to form a loop of open strings,
one should make the worldsheet time $\tau$ periodic, but this is
incompatible with the light-cone gauge $X^+ = x_0^+ + p^+ \tau$ 
unless $p^+=0$.

On the other hand, it is well known that a non-trivial contribution
to the vacuum energy is obtained when the light-cone coordinate $X^+$
is compact of radius $R$. Indeed, the integral over $p^-$ in \eqref{coleman}
is turned into a discrete sum on integers $p^- = n/R$, or after Poisson
resummation into a periodic delta distribution with support at 
$p^+ = w R/t$, $w\in \Zint$. One-loop vacuum energy with a loop modulus of
$2 \pi t$ writes
\be
\label{ccoleman}
A_{op} = \frac12 V_{+-} \sum_{w=-\infty}^{\infty} \int_0^{\infty} \frac{dt}{t^2} 
\left. \Tr \exp (i 2\pi m^2 t) \right\rvert_{p^+=R w /t}
\ee 
The quantization of $p^+$ is now in agreement with the fact that $X^+$ should
be a periodic function of $\tau$ of period $2\pi t$, modulo the identification 
$X^+ \sim X^+ + 2\pi w R$.
For open strings, $m^2 = p^+ {\cal H}_{op}$ and after a Wick rotation, one obtains
\be
\begin{split}
 A_{op} & = -\frac{i}2 V_{+-}\sum_{w=-\infty}^{\infty} \int_0^{\infty} \frac{dt}{t^2} 
\left. \Tr \exp\left( - 2\pi p^+ {\cal H}_{op} t \right) \right\rvert_{p^+=R w /t} \\
 & = -\frac{i}2 V_{+-}\sum_{w=-\infty}^{\infty} \int_0^{\infty} \frac{dt}{t^2} Z_{op} (t)
\end{split}
\ee

As usual, the open string vacuum amplitude should be reexpressible in
the closed string channel as the propagator of closed strings between
two boundary states describing the two D-branes with the electromagnetic
flux on them. Since our D-branes have Neumann boundary condition along the 
light-cone coordinate $X^+$, it is not possible to use the standard
light-cone gauge to quantize the closed strings. Instead, one may use
$X^+ = w R \sigma$, which follows directly from the open-string 
light-cone gauge $X^+ = p^+ \tau$ after exchanging and rescaling the
worldsheet coordinates, $\tau \to t \sigma$. The integer $w$ is now
interpreted as the winding number of the closed string around the
compact coordinate $X^+$. The closed string amplitude therefore reads
\be
A_{cl} = \sum_i \langle B_1 \rvert \frac{1}{p^+ p^- + m_i^2} \lvert B_0 \rangle
\ee
where we used the fact that the boundary states $B_0$ and $B_1$ are
annihilated by $p^+$ and $p^-$.
For closed strings propagation with Schwinger time $\pi \hat{t}$
\be
\begin{split}
A_{cl} &= \sum_{w=-\infty}^{\infty} \int_0^{\infty} d\hat{t} 
 \langle B_1(w) | e^{-\pi \hat{t} H_{cl} } \lvert B_0(w)  \rangle   \label{clo}  \\
 &=  \sum_{w=-\infty}^{\infty} \int_0^{\infty} d\hat{t}~V_{+-}~Z_{cl} (\hat{t})
\end{split}
\ee
The equality between \eqref{ccoleman}
and \eqref{clo} for D-branes in flat space now follows from
the usual modular properties of open/closed string partition functions.

We now return to the case of charged open strings in a constant quadrupolar
electromagnetic wave, $\Phi_a = (h_a)_{ij} x^i x^j$ and compute
the one-loop vacuum energy in the case where the light-cone coordinate $X^+$
is compact. Then, boundary conditions for open and closed strings involve parameter $e$ and
$\hat{e}$ which are
\be e_a = \pi p^+ h_a = \pi w R h_a /t\ , \qquad \hat{e}_a = \pi w R h_a = \pi p^+ h_a t, 
\label{hate}\ee
and thus obey the duality relation $\hat{e}_a = e_a t$
(in fact, there is a $i$ coming from the Wick rotation, see \eqref{rescaling}).
Starting with the open string channel, the contribution of the coordinate $X$ with $e_0^X = e_0$
and $e_1^X = e_1$ on a cylinder of modulus 
$q=e^{-2\pi t}$ is given, with due care paid to the zero-mode, by
\be
\label{partfunc} 
Z_{op}(t,e_0,e_1) 
= q^{E_X}  \prod_{n=1}^\infty \left(1 - q^{\omega_n} \right)^{-1} \times
\begin{cases} 
 \left(1 - q^{\omega_0} \right)^{-1}  & \text{if } D > 0 \\ 
\frac{1}{2 t} \sqrt{\frac{(2-3e_0+e_0^2)}{3(1-e_0)D}}
& \text{if } D\to 0 \\
 \left(1 - q^{ik_0} \right)^{-1}  & \text{if } D < 0 
\end{cases} 
\ee
In the zero field limit, one retrieves the correct behaviour
\be Z_{op}(t,e_0,e_1) \sim \frac1{\sqrt{e_0-e_1}t~\eta(t)} 
\sim \frac1{\sqrt{ht}~\eta(t)} \sim \frac{V}{\sqrt{t}~\eta(t)} \ee 
where V is the volume accessible to 0-modes and $\eta(t) =
q^{\frac1{24}} \prod_{n=1}^\infty \left(1 - q^n \right)$ is the Dedekind eta
function.
The one-loop open string amplitude is obtained by putting together the
contributions of the $x$ and $y$ directions as well as the remaining 22
transverse coordinates $x^i$, and reads
\be
A_{op} = -\frac{i}2 V_{+-} \sum_{w=-\infty}^{\infty}
\int_0^{+\infty} \frac{d t}{t^2} \, {\cal V}_{22} (4 \pi^2 t \eta^2(t))^{-11}
Z_{op}(t,e_0, e_1) Z_{op}(t,-e_0 ,-e_1) 
\label{1loopo}
\ee
where ${\cal V}_{22}$ is the volume of the 22 free transverse directions.
For $D<0$, the partition function picks an imaginary part, due to the
unstable mode $k_0$. It would be interesting to extract the production
rate of macroscopic strings, in analogy with the stringy computation of 
Schwinger pair production in \cite{Bachas:bh}.

Now we turn to the closed string channel. 
As explained above, it is
convenient to use a non-standard light-cone gauge $X^+(\tau,\sigma) =Rw \sigma$,
compatible with the Neumann boundary condition on $X^+$.
Focusing on a single transverse boson $X$, the boundary 
condition \eqref{bcquad} at $\sigma=0$
becomes a condition at one cap $\tau=0$ of the cylinder, 
\be
\label{bcdual}
\pt X + \frac{\hat{e}}{\pi} X = 0 \quad\mbox{at}\  \tau=0
\ee
with $\hat{e}/\pi$ given in \eqref{hate}. 
The coordinate $X$ still satisfies the free field equation 
$(\p_\tau^2 -\p_\sigma^2)X=0$ in the bulk, and can be expanded
in the usual free closed string modes,
\be 
X_{cl} = x_0 + p_0 \tau + i\sum_{n \neq 0}  \frac{1}{n} 
\left(\alpha_n e^{-in (\tau+\sigma)}  
+\tilde{\alpha}_n e^{-in (\tau - \sigma)} \right)
\ee
The condition \eqref{bcdual} can now be written as a condition on a boundary
state $|B(\hat{e})\rangle$ in the closed string Hilbert space,
\bse
\begin{gather}
\left[\left(1 + \frac{i\hat{e}}{\pi n} \right) \alpha_n + \left( 1-\frac{i\hat{e}}{\pi n} \right) 
\tilde{\alpha}_{-n} \right] \left\vert B(\hat{e}) \right\rangle = 0 \\
\left(p_0 + \frac{\hat{e}}{\pi} x_0 \right)\left\vert B(\hat{e}) \right\rangle = 0 \label{zeroconstr}
\end{gather}
\ese
which can be solved as usual by coherent state techniques,
\be
\vert B(\hat{e})\rangle = {\cal N}(\hat{e})\, e^{i\frac{\pi p_0^2}{2 \hat{e}}} 
\exp \left( \sum_{n=1}^\infty -\frac{1}{n} \frac{i\pi n+\hat{e}}{i\pi n-\hat{e}}
\alpha_{-n} \tilde{\alpha}_{-n}\right) 
\vert 0,\tilde{0} \rangle 
\ee
where 
\be {\cal N}(\hat{e})  = \frac1{\sqrt{2\sqrt{2} \sinh(\hat{e})}}. \tag{\ref{normbndry}} \ee 
is fixed by open-closed duality as explained in Appendix \ref{opcldual}.
Here we used a momentum representation to solve the zero-mode constraint
\eqref{zeroconstr}. The contribution of the coordinate $X$ with $\hat{e}_0^X =\hat{e}_0$
and $\hat{e}_1^X= \hat{e}_1$ to the
one-loop amplitude in the closed string channel is now simply obtained
as
\be
\label{matrixelem}
Z_{cl} = \int dp_0 
\langle B(\hat{e}_1) \rvert e^{-\pi \hat{t} p^+ {\cal H}^{cl}} 
\lvert B(\hat{e}_0) \rangle 
\ee
where we included a summation over the closed string zero-mode $p_0$ along
the direction $x$. The free closed string Hamiltonian reads as usual
\be
p^+ {\cal H}^{cl}= \frac 12 p_0^2 + \sum_{n=1}^{\infty} (\alpha_{-n} \alpha_n  + 
\tilde\alpha_{-n} \tilde\alpha_n) - \frac1{24}
\ee
The matrix element \eqref{matrixelem} can now be evaluated using the identity
\be
\left< 0,\tilde{0} \right\vert e^{\lambda_1^* \alpha \tilde{\alpha}} 
e^{- \pi \hat{t} (\alpha^\dag \alpha +  \tilde{\alpha}^\dag \tilde{\alpha})}
e^{\lambda_0 \alpha^\dag \tilde{\alpha}^\dag}
\left\vert 0,\tilde{0} \right>
=\frac{1}{1-\lambda_1^* \lambda_0 e^{-2\pi \hat{t}}} 
\ee
where we assumed the commutation relations $[\alpha,\alpha^\dagger]=
[\tilde\alpha,\tilde\alpha^\dagger]=1$. The integral on the zero-mode
$p_0$ is Gaussian. Altogether, we obtain, keeping in mind that the 
commutation relations for the 
$\alpha_n$'s and the $\tilde{\alpha}_n$'s are $[\alpha_n, \alpha^\dag_n] = n$ and
$[\tilde{\alpha}_n, \tilde{\alpha}^\dag_n] = n$ :
\be
\label{amplclosed}
Z_{cl}(\hat{t},\hat{e}_0,\hat{e}_1) =  {\cal N}(\hat{e}_0)  {\cal N}(\hat{e}_1)
 \sqrt{\frac{2}{\hat{t} + i \left(\frac1{\hat{e}_1} - \frac1{\hat{e}_0} \right)}} \hat{q}^{-\frac1{24}}
 \prod_{n=1}^{\infty} 
\left(1-\frac{i\pi n+\hat{e}_0}{i\pi n-\hat{e}_0} \frac{i\pi n-\hat{e}_1}{i\pi n+\hat{e}_1} 
\hat{q}^n \right)^{-1} 
\ee
where $\hat{q} = e^{-2\pi \hat{t}}$. 
In the zero field limit, one obtains correctly
\be Z_{cl}(\hat{t},\hat{e}_0,\hat{e}_1) \sim \frac1{\sqrt{\hat{e}_0-\hat{e}_1} \eta(\hat{t})}
\sim \frac1{\sqrt{h}~\eta(\hat{t})} \sim \frac{V}{\eta(\hat{t})} \ee
One therefore obtains the one-loop amplitude in the closed string channel,
\be 
A_{cl} = \sum_{w=-\infty}^{\infty}
\int_0^{+\infty} d \hat{t} \, {\cal V}_{22}\, \eta(\hat{t})^{-22}
Z_{cl}(\hat{t},\hat{e}_0, \hat{e}_1) 
Z_{cl}(\hat{t},-\hat{e}_0, -\hat{e}_1)
\ee
In order to establish the equality with the amplitude in the 
open string channel, we need to study the transformation law of
$Z$ under $\hat{t}\to 1/\hat{t}$. A heuristic derivation can be performed
as follows, generalizing techniques introduced in \cite{Arutyunov:2001nz} for
computing the vacuum energy -- a full-fledged demonstration is
provided in Appendix \ref{opcldual}, building instead on methods introduced
in \cite{Gaberdiel:2002hh}. The logarithm of the open string partition
function, disregarding vacuum energy contributions, can be written
as the contour integral
\be
\log Z_{op} = \frac{1}{2\pi} \int_C dz  (\log \Phi_{cl}) 
\frac{d\log \Phi_{op}}{dz}
\ee
where $d\log \Phi_o/dz$ has single poles with unit residue 
at the open string eigenmodes, while $\log \Phi_c$ provides 
the correct contribution of that eigenmode to the free energy:
\be
\Phi_{op}(z) = 1 - e^{-2\pi i z}  \frac{i\pi z + e_0}{i\pi z - e_0}
\frac{i\pi z - e_1}{i\pi z + e_1} \ ,\quad
\Phi_{cl}(z) = 1 - e^{-2\pi t z}
\ee
The contour $C$ follows the positive real axis from below 
from $+\infty-i\epsilon$ down to $z=\epsilon<\om_0$, and then
back to $+\infty+i\epsilon$ from above. We may now unfold the contour
to the imaginary axis $C'=]-i\infty+\epsilon,-i\infty+\epsilon[$, and
integrate by part: 
\be
\log Z_{op} = \frac{1}{2\pi} \int_C' dz  (\log \Phi_o) 
\frac{d\log \Phi_c}{dz}
\ee
The term $d\log \Phi_c/dz$ now has single poles at $z=i n/t$, corresponding
to the closed string oscillators. The factor $\log \Phi_o$ provides the
correct contribution to the logarithm of the closed string amplitude,
upon redefining $\hat{e}_0 = -it e_0, \hat{e}_1 = -it e_1$. More rigorously,
we show in Appendix \ref{opcldual} the relation
\be
\label{ocd}
Z_{op}(t,e_0, e_1)
Z_{op}(t,-e_0, -e_1) =  
Z_{cl}(\hat{t},\hat{e}_0, \hat{e}_1)
Z_{cl}(\hat{t},-\hat{e}_0, -\hat{e}_1)
\ee
This gives a new example of open-closed duality in a non-conformal
setting, beyond the computation in \cite{Bergman:2002hv}.

Finally, let us comment on the issue of fermionic strings in such 
backgrounds. The coupling of the Ramond-Neveu-Schwarz superstring
to an electromagnetic background can be written as $\oint dz
\bar \psi^\mu \psi^\nu F_{\mu\nu}$. In the light-gauge, $\psi^+=0$,
hence the coupling vanishes for a null electromagnetic field
$F_{i+}$. This implies that the fermions remain quantized 
in integer or half-integer modes just as in flat space. While
this seems to clash with worldsheet supersymmetry, the latter
is broken by the choice of light-cone gauge\footnote{We thank A. Tseytlin 
for emphasizing this to us}. This is another
difference with tachyon condensation, where worldsheet supersymmetry
requires to introduce a non-local interaction for the fermions
\cite{Kutasov:2000aq}. The supersymmetries unbroken by the 
electromagnetic background correspond to kinematical symmetries,
which commute with the Hamiltonian and do not mix worldsheet
bosons with fermions.

\section{Open strings in time-dependent electric field}
We now consider an open string moving in a time-dependent quadrupolar 
electric field. Due to the translational symmetry along $x^-$, there
is as usual no production of strings at zero string coupling. On the
other hand, a single string will generically get excited as it passes
through the electromagnetic wave. This issue was studied in detail
in \cite{hs} for closed strings in gravitational waves. In this
section we adapt their analysis to the case of open strings.

\subsection{Mode production and Bogolioubov transformation}

We assume a ``sandwich wave'' configuration, i.e. that the profile
$h_a(x^+)$ has a compact support in $x^+$, and concentrate on 
a single direction $x$. At early times, we may expand the embedding 
coordinate $X(\tau,\sigma)$ into free open string modes,
\be
X(\tau,\sigma) = x_0 + p_0 \tau + i \sum_{n\neq 0}
\frac{2}{n} a_n \cos(n \sigma) e^{-i n \tau}\ ,\quad \tau \to -\infty
\ee
A similar mode expansion may be carried out at late times by
replacing $(x_0,p_0,a_n)$ by $(y_0,q_0,b_n)$. Since the
perturbation is linear, the two sets of modes will be related by
a linear Bogolioubov transformation,
\be 
\label{lbog}
\begin{pmatrix} y_0 \\ q_0 \\ b_m \end{pmatrix} = 
\begin{pmatrix} \alpha & \beta & A_n \\
\gamma & \delta & B_n \\
\tilde A_m &  \tilde B_m & 
B_{mn} \end{pmatrix}
\begin{pmatrix} x_0 \\ p_0 \\ a_n \end{pmatrix} 
\ee
Since the equation of motion \eqref{diffdiff} has real coefficients, this
matrix satisfies the reality properties
\be
(\alpha,\beta,\gamma,\delta)\in \Real^4\ ,\quad
(A,B,\tilde A,\tilde B)^*_m = 
(A,B,\tilde A,\tilde B)_{-m}\ ,\quad
B_{mn}^* = B_{-m,-n}
\ee
in addition to preserving the symplectic form 
$-i dx_0 \wedge dp_0 + \sum_{m> 0} da_m \wedge da_{-m}/m$.
In particular, the off-diagonal components $B_{mn}$ with $m>0$ and
$n<0$ relate creation and annihilation operators at
$\tau\to\pm\infty$, hence encode mode creation. This can be
seen by evaluating the occupation number of the $n$-th mode
at late times,
\be \label{occupancy}
\begin{split}
\langle 0_{in} | b_{-m} b_m | 0_{in} \rangle
=& \sum_{n \neq 0} |B_{m,-n}|^2 \\
&+ \int_{-\infty}^\infty  dx \psi^*(x) 
\left( \tilde A_{-m} x + \tilde B_{-m} \frac{d}{idx} \right)
\left( \tilde A_{m} x + \tilde B_{m} \frac{d}{idx} \right)
\psi(x)
\end{split}
\ee
where $\psi(x)$ is the quantum wave function of the open string
zero-mode $x_0$ at early times. 

While the Bogolioubov transformation \eqref{lbog} suffices to describe
the effect of the sandwich wave on the string modes, it is important to
observe that it fails to account for an extra degree of freedom which
can be attributed to the background itself. Indeed, the zero-mode 
$x_0$ of the string is really the sum $f_0+g_0$ of zero-modes for
the left and right-moving components, respectively. The difference
$f_0-g_0$ usually has no physical significance, since it can be
interpreted as a constant $U(1)$ gauge potential $A=(f_0-g_0)dx$
along the direction $x$, which can be gauged away -- indeed,
under T-duality $f_0-g_0$ maps to the position of the D-brane along
the transverse direction $x$, which is itself T-dual to the 
gauge potential $A_x$. In a time-dependent situation however, the
variation of $f_0-g_0$ between $\tau=-\infty$ and $\tau=+\infty$
has a gauge-invariant meaning, namely the integral of the electric field
$A_x(+\infty)-A_x(-\infty)= \int F_{+x} dx^+$
or equivalently the shift in transverse position of the T-dual D-brane.
This change of the electromagnetic background induced by the propagation
of strings is possibly the simplest instance of back-reaction
in string theory.


\subsection{Born approximation}

Let us now consider the regime in which the electric field can be 
treated as a perturbation of the free open string case. This is
in particular the case of an highly excited state of the string.
We assume that the string starts out in a mode $a_n$, and expand
the left and right movers to first order in $h$,
\be
\begin{pmatrix} f \\ g \end{pmatrix}
= \frac{i}{n} a_n 
\begin{pmatrix} 1 \\ 1 \end{pmatrix} e^{-i n \tau}
+
\begin{pmatrix} \delta f \\ \delta g \end{pmatrix}
\ee
where $\delta f$ and $\delta g$ are assumed to be of order $h$. The
equation of motion \eqref{diffdiff} can now be rewritten as a 
linear differential difference equation with source,
\be
\begin{pmatrix} D & -D \\
DT^2 & -D 
\end{pmatrix}
\begin{pmatrix} \delta f \\ \delta g
\end{pmatrix}
= - \frac{2i a_n}{n\pi}  \begin{pmatrix} e_0 \\
(Te_1) \end{pmatrix} e^{-i n \tau}
\label{eqpert}
\ee
where $T$ is the shift operator by $\pi$ as above.
The retarded Green function for this linear system is easily computed
by Fourier analysis
\be
G(\tau)= \int_{-\infty}^{\infty} \frac{d\omega}{2\pi i \omega}
\begin{pmatrix} 1 & -1 \\ e^{-2\pi i \omega} & -1 \end{pmatrix}
\frac{e^{-i\om \tau}}{e^{-2\pi i \omega} -1}
\ee
and can be easily evaluated, with the result\footnote{A closely
related computation was performed independently in \cite{Bachas}.}
\be
G(\tau)= -\frac{1}{2\pi} 
\begin{pmatrix} \tau-\pi & \pi - \tau \\ \tau+\pi & \pi -\tau \end{pmatrix}
+\sum_{m\in\Zint^*}
\begin{pmatrix} 1 & -1 \\ 1 & -1 \end{pmatrix}
\frac{e^{-im \tau}}{2\pi i m}
\ee
Note that $G$ is the solution of the Green equation with a minus sign, $ {\cal D} G = \delta$, 
with $\cal D$, the  matrix of differential operators on the \lhs of \eqref{eqpert}.
One therefore obtains the solution after the electromagnetic 
wave has passed,
\be
\begin{split}
\begin{pmatrix} f \\ g \end{pmatrix}
= \frac{i}{n} a_n e^{-i n \tau}
\begin{pmatrix} 1 \\ 1 \end{pmatrix}
&+ 
\frac{i a_n}{n \pi} \int_{-\infty}^\infty 
\begin{pmatrix} (Te_1)-e_0 \\ e_0+(Te_1) \end{pmatrix} (p^+ \tau')
e^{-in\tau'} d\tau' \\
&+ \frac{i a_n}{n \pi^2} \int_{-\infty}^\infty 
 (\tau-\tau') (e_0-(Te_1))(p^+\tau') e^{-in\tau'} d\tau'
\begin{pmatrix} 1 \\ 1 \end{pmatrix} \\
&-\frac{ia_n}{\pi n} \sum_{m\in\Zint^*} \left( \int_{-\infty}^\infty 
 (e_0-(Te_1))(p^+\tau')
 \frac{1}{\pi i m} e^{-i(n-m)\tau'} d\tau' \right)
e^{-im \tau} \begin{pmatrix} 1 \\ 1 \end{pmatrix} .
\end{split}
\ee
One deduces the matrix element of the Bogolioubov transformation
\bse
\begin{align}
B_{mn} &= \delta_{mn} + 
\frac{i}{\pi^2 n}  
\int_{-\infty}^\infty  (e_0-(Te_1))(p^+\tau) e^{-i(n-m)\tau} d\tau\\
A_n &=  -\frac{2i}{n\pi^2} \int_{-\infty}^\infty  
 \left(\tau (e_0-(Te_1))(p^+\tau) - \pi(Te_1)(p^+ \tau)\right)
e^{-in\tau} d\tau\ , \\
B_n &= 
 \frac{2i}{n\pi^2}  \int_{-\infty}^\infty  (e_0-(Te_1))(p^+\tau) e^{-in\tau} d\tau\ , 
\end{align}
\ese
The other matrix elements can be obtained by starting with a zero
mode $x_0 + p_0 \tau$ at $\tau\to -\infty$ , yielding
\bse
\begin{align}
\alpha &= 1 - \frac{1}{\pi^2} \int_{-\infty}^\infty  
\left(\tau (e_0-(Te_1))(p^+\tau)
  - \pi(Te_1)(p^+ \tau)\right)d\tau\ , \\
\beta &= - \frac{1}{\pi^2} \int_{-\infty}^\infty  
\left(\tau^2 (e_0-(Te_1))(p^+\tau)
-\pi(\pi +2\tau) (Te_1)(p^+\tau) \right) d\tau\ , \\
\gamma &= \frac1{\pi^2} \int_{-\infty}^\infty  (e_0-(Te_1))(p^+\tau) 
d\tau\ , \\
\delta &= 1 + \frac1{\pi^2} \int_{-\infty}^\infty 
 \left( \tau (e_0-(Te_1))(p^+\tau) - \pi(Te_1)(p^+\tau)\right) d\tau \\
\tilde{A}_m &= \frac1{2\pi^2} \int_{-\infty}^\infty 
 (e_0-(Te_1))(p^+\tau) e^{im\tau} d\tau\ , \\
\tilde{B}_m &=  \frac1{2\pi^2} \int_{-\infty}^\infty 
 \left(\tau(e_0-(Te_1))(p^+\tau) - \pi(Te_1)(p^+\tau) 
\right) e^{im\tau} d\tau\ .
\end{align}
\ese
and using the symplectic properties of the Bogolioubov transformation.
One may in fact keep trap of the splitting of the zero-mode $x_0=f_0+g_0$
between the left- and right-movers, and find the variation of $f_0$
and $g_0$ separately between $\tau=-\infty$ and $\tau=+\infty$:
\be 
\delta f_0-\delta g_0 
= - \frac{1}{\pi}
\int_{-\infty}^{\infty} e_0 (p^+ \tau) X(\sigma=0,\tau) d\tau 
 \label{deltapos}
\ee
As discussed in the previous section, this variation amounts to a correction
to the null electric background $F_{x+}$, and 
hence to a backreaction of the string on the electric background.

As in the case of closed strings in time-dependent gravitational plane
waves \cite{hs}, it is of interest to ask if an open string can smoothly
pass through a singular wave profile. This can be answered by computing
the occupation number of the oscillator $a_n$ after the wave has
passed from \eqref{occupancy}. If the wave profile $h^a$
is a smooth ($C^{\infty}$) function with compact support in $x^+$, 
its Fourier transform $\hat h_a$ decreases faster than any power, and 
therefore the total excitation number is finite. In the case of an
impulsive wave, where $h_a$ has a delta function
singularity, the Fourier transform $\hat h_a(\omega)$ goes 
to a constant at $|\omega|\to\infty$:
the total energy of the outgoing string $\sum n \langle  N_n \rangle$ 
is infinite, implying that the string is torn apart as it goes through
the singularity. For a shock wave with profile $h^a$ proportional 
to the Heaviside step function, the total energy is instead finite,
implying a smooth propagation. For the case of a ``conformal singularity''
$h^a(x^+)\sim 1/x_+^2$ discussed recently in the context of closed
strings \cite{Papadopoulos:2002bg}, 
the Fourier transform $\hat h^a$ diverges linearly
and therefore the total excitation number itself is infinite.

\subsection{Adiabatic approximation}
We now consider the motion of open strings in an adiabatically
varying electric field. The transverse coordinate
$X$ can still be decomposed into left-movers and right-movers.
One may eliminate the latter to get an equation for $f(\tau)$ only,
\be
0=(1-T^2) \ddot{f} 
-\left[ p^+ \frac{(T\dot{e}_1) - \dot{e}_0}{(Te_1) - e_0} (1-T^2) 
+  \frac1\pi((Te_1) -e_0) (1+T^2)\right]\dot{f} 
\ee
\be \nn
+ \left[\frac{p^+}{\pi} (\dot{e}_0 - (T\dot{e}_1) T^2)
-\frac{p^+}{\pi} 
\frac{ (T\dot{e}_1)-\dot{e}_0}{(Te_1) - e_0} (e_0-(Te_1) T^2) 
-\frac1{\pi^2} e_0 (Te_1)(1-T^2)\right]f
\ee
where the dot denotes differentiation with respect to $\tau$,
$T$ is the shift operator by $\pi$ and in $(Te_1)$, 
$T$ acts only on $e_1$. The case $e_0 = (Te_1) \stackrel{\text{\scriptsize def}}{=} e$ 
gives a more simple equation
\be (1-T^2) \dot{f} + e (1-T^2) f = 0 \ee 
which can be solved exaclty as an infinite sum of shifted solutions of $\dot{y} + e y =0$.
However, the expression are hardly tractable and we will not use them.

In line with the standard WKB approach, 
we assume that the phase of $f$ varies much faster than its modulus.
We thus write $f = s \exp{i\gamma}$, rescale $\gamma\to
\gamma/\epsilon$ and time $\tau\to \tau/\epsilon$.
Note that shift operator is also rescaled, $Tx(\tau) = x(\tau + \e \pi)$.  
At leading order in $\e$, we recover the standard dispersion relation
\be
\label{stringadiabzero} 
((\pi\dot{\gamma}_0)^2 + e_0 e_1) \sin(\pi \dot{\gamma}_0) - 
(e_1- e_0) \pi\dot{\gamma}_0 \cos(\pi \dot{\gamma}_0) =0 
\ee
We choose for $\dot{\gamma}_0(\tau)$ one of the branches 
in \eqref{stringadiabzero}, which we assume to stay real at all times.
This is for instance the case of the excited states $\omega_{n>1}$.
To  next-to-leading order in $\epsilon$, we obtain the variation in
amplitude (as well as a correction to the phase),
\be
s_0(\tau) = s_0(0) \exp \left(-i\pi \left(\dot{\gamma_0}(\tau) - \dot{\gamma_0}(0) \right)
 - \int_0^\tau \frac{{\cal B} (\tau')}{{\cal A}(\tau')} d \tau'\right) ,
\ee
where
\bse
\begin{align}
{\cal A} &= -\pi \dot{\gamma}_0^4 - \frac{1}{\pi} \left[ e_0 (1- e_0) - e_1 (1+e_1) \right]
 \dot{\gamma}_0^2 + \frac{e_0 e_1}{\pi^3} (e_1 - e_0 - e_0 e_1), \\
{\cal B} &= -2\pi \dot{\gamma}_0^3 \ddot{\gamma}_0 +i p^+ \dot{e}_1 \dot{\gamma}_0^3 + 
\frac{p^+}{\pi} (\dot{e}_1 - \dot{e}_0) \dot{\gamma}_0^2 + \notag \\ 
 & \phantom{= -} + \frac{1}{\pi} \left[ e_0 (1- e_0) - e_1 (1+e_1) \right] \dot{\gamma}_0 \ddot{\gamma}_0 +
\frac{i p^+}{\pi^2} (\dot{e}_0 e_1 - e_0 \dot{e}_1 +ie^2_0 \dot{e}_1) .
\end{align}
\ese
This result is valid as long as $\dot e\ll e^{3/2}$ and that no
phase transition occurs. If a classically disallowed region is
encountered, one should take into account the reflected wave 
using the standard WKB matching techniques.
As in the dipole case, it does not seem possible to eliminate the
phase altogether and get a deformed classical equation as in \eqref{lewis}
valid outside the adiabatic regime.

\subsection{Sudden approximation - open string in a Paul trap}
Finally, we consider the regime opposite to the adiabatic
approximation, where the perturbation is switched on and off
by a Heaviside function of time. We have in mind the situation
of the Paul trap, where the electric gradient is modulated 
by a square profile,
\bea
\Phi_a&=& \frac12 h_a x^2 \ ,\quad x^+\in [nT,(n+\frac12 T)[\\
\Phi_a&=& \frac12 h'_a x^2 \ ,\quad x^+\in [(n+\frac12 T),(n+1)T[\ .
\eea
The embedding coordinate $X$ can be expanded in modes $a_n$
or $a'_n$ as in \eqref{modexpansion} and \eqref{genmode}
in either of the two intervals. The two
sets of modes can be related by matching $X(\sigma,\tau)$ and
its derivative at $x^+=0$ and $x^+=T/2$  using the orthogonality relation
\eqref{orthogonality}. At $x^+=0$, we find
\be
a'_n = S_{nm}(h',h) a_m
\ee
where 
\be
\begin{split}
S_{nm}(h',h) &= i \om'_n
\int_0^\pi \left(X_{-n,h'} \p_\tau X_{m,h} - X_{m,h} \p_\tau X_{-n,h'} \right) d\sigma \\
&= i\om'_n {\cal N}_{n,h'}{\cal N}_{m,h} \left[
\frac{\om'_n+\om_m}{\om'_n \om_m(\om'_n-\om_m)}\left( 1 - \sqrt{\frac{\Omega_m^0}{{\Omega'}_n^0}}
\left(\sqrt{\frac{{\Omega'}_n^1}{\Omega_m^1}}-\sqrt{\frac{\Omega_m^1}{{\Omega'}_n^1}}\right)
- \frac{\Omega_m^0}{{\Omega'}_n^0} \right) \right. \\
&\qquad \qquad + \left. \frac{1}{\om'_n \om_m} \left( 
\Omega_m^0 - \frac{1}{{\Omega'}_n^0} + \sqrt{\frac{\Omega_m^0}{{\Omega'}_n^0}}
\left(\sqrt{\frac{1}{\Omega_m^1 {\Omega'}_n^1}} - \sqrt{\Omega_m^1 {\Omega'}_n^1} \right)\right) \right]\ ,
\end{split}
\ee
with $\Omega_n^i = \frac{i\pi \om_n - e_i}{i\pi \om_n +e_i}$ and the corresponding definition with 
primes on $e_i$ for ${\Omega'}_n^i$.
When $h' = h$, $S_{nm}(h,h) = \delta_{nm}$ as it should.
One therefore obtains the transition matrix from one period to the
next,
\be
T_{nm} =  \sum_{p\in\Zint}   e^{i \frac{T}{2p^+}\omega_n}   S_{np}(h,h') 
e^{i  \frac{T}{2p^+}\omega'_p}  S_{pm}(h',h)
\ee
The issue of stability of open strings in the Paul trap is therefore
reduced to the determining the modulus of the eigenvalues
of the transfer matrix $T$. We expect a behaviour similar to the
elastic dipole studied in \ref{Paul}.

\acknowledgments
It is a pleasure to thank C. Bachas, G. D'Appollonio, M. Gutperle,
B. Stefanski and A. Tseytlin for valuable discussions and suggestions, 
and 
especially C. Bachas and A. Tseytlin for critical comments on the manuscript. 

\vskip 5mm
\noindent {\bf Historical note}: After the first version of this paper was
released on the arXiv, we became aware of \cite{Arutyunov:2001nz,
Bardakci:2001ck} where similar boundary conditions are discussed
in the context of tachyon condensation at one-loop, with partial
overlap with the mathematical results in Section 4. Section 4.3 on
fermionic strings and 5.2 on neutral strings in time-dependent
plane waves have been retracted in the present version, as
it was realized that worldsheet supersymmetry need not be present
on the light-cone, and that the appropriate condition $e_i=T^2 e_i$
that leads to a simplification of the dynamics also renders the question
of the Bogolioubov transformation moot. Finally, Section 4.1.3
on the one-loop string amplitude in the first release missed the fact,
made apparent to us by the recent paper \cite{Hikida:2003bq},
that the light-cone needs to be compact in order to produce a non-trivial
result; we have clarified this point in Section 4.3. As a reward to 
the reader for bearing with these errands, this version provides added
value in the form of an observation on an hidden degree of freedom
of the background (Section 5.1 and 5.2), and a derivation of the
open-closed duality formula (Appendix \ref{opcldual}).

\appendix
\vskip 5mm
\centerline{\bf \Large Appendices}
\vskip 5mm

\section{Rigorous stability analysis}
\label{proof}
In this Appendix, we establish the criteria for stability of open strings in a purely
electromagnetic quadrupolar trap on rigorous
ground, following techniques explained in \cite{hale}. The question 
to address is under which condition the characteristic equation 
$\Delta(z)=P(z,e^z)$ has all its zeros in the left half plane $\Re(z)\leq 0$,
where $P(x,y)$ is a polynomial in $x,y$. For neutral differential
difference equations, as is the case of interest in this paper, 
$P(x,y)$ admits a principal term, i.e. contains a 
monomial $a_{rs} x^r y^s$ such that all the other terms $a_{mn} x^m y^n$
have either $r>m,s>n$, or $r=m,s>n$, or $r>m,s=n$. The analysis
is then based on the following theorem (see \cite{hale}
for details):

\begin{theorem}{\bf Theorem 1} {\it Pontryagin (\cite{hale}, A.3)}
Let $\Delta(iy)=F(y)+iG(y)$ be the real and imaginary parts of $\Delta(iy)$
for $y\in \Real$. If all zeros of $\Delta(z)$ have negative real parts,
then the zeros of $F(y)$ and $G(y)$ are real, simple, alternate and
\be
\label{posc}
G'(y) F(y) - G(y) F'(y) \geq 0
\ee
for all $y\in \Real$. Conversely, all zeros of $\Delta(z)$ will be in the left
half plane provided that either of the following conditions is satisfied:
\begin{enumerate}
\item{(i)} All the zeros of $F(y)$ and $G(y)$ are  real, simple, alternate 
and \eqref{posc} is satisfied for at least one $y\in \Real$.
\item{(ii)} All the zeros of $F(y)$ are  real, and for each zero,
\eqref{posc} is satisfied.
\item{(iii)} All the zeros of $G(y)$ are  real, and for each zero,
\eqref{posc} is satisfied.
\end{enumerate}
\end{theorem}
In order to establish condition (ii) or (iii), one rewrite $F$ or $G$ as
$f(y,\cos(y),\sin(y))$, which one decomposes as
\be
\label{decref}
f(z,u,v)=\sum_{m=0}^r \sum_{n=0}^s z^m \phi_m^{(n)}(u,v)
\ee
where $\phi_m^{(n)}$ is a homogenous polynomial of degree $n$ in $(u,v)$.
The principal term in $f$ is defined as the term $z^r \phi_r^{(s)}(u,v)$
such that either $r>m,s>n$, or $r=m,s>n$, or $r>m, s=n$ in \eqref{decref}.
We then have the theorem
\begin{theorem}{\bf Theorem 2} \label{th2} {\it Pontryagin (\cite{hale}, A.4)}
If $\epsilon$ is such that $\phi_r^{(s)}(\cos(iy+\epsilon),\sin(iy+\epsilon))$
is non vanishing for all $y\in\Real$, then, for sufficiently large integers
$k$ the function $F(z)=f(z,\cos(z),\sin(z))$ has exactly $4ks+r$ zeros
in the strip $-2k\pi+\epsilon\leq \Re(z)\leq 2k\pi+\epsilon$. Consequently,
the function $F(z)$ will have only real roots iff, for sufficiently 
large integers $k$, it has exactly  $4ks+r$ real zeros
in the interval $I_k=[-2k\pi+\epsilon,2k\pi+\epsilon]$.
\end{theorem}
We now have the tools to investigate the stability of the linear system
\eqref{diffdiff}. Identifying $z=2\pi i\omega$, the characteristic equation
can be written as
\be
\Delta(z)=(-z^2+4 e_0 e_1)(e^z-1)-2(e_1-e_0)z(e^z+1)
\ee
Consequently, the real and imaginary parts of $\Delta(iy)$ read
\bea
F(y)&=&(y^2+ e_0 e_1)(\cos y-1)-2(e_0-e_1)y\sin y\\
G(y)&=&(y^2+4 e_0 e_1)\sin y +2(e_0-e_1)y(1+\cos y)
\eea
The condition \eqref{posc} is satisfied trivially, since
\be
F'G-FG'=2\left[ (y^2+4 e_0 e_1)\sin(y/2)+2(e_0-e_1)y\cos(y/2)
 \right]^2
\ee
as a consequence of the fact that the argument of $\Delta(iy)$ is $y/2$.
The condition of stability is therefore that $G(y)$ has only real roots.
$G(y)$ can be written as $f(z,\cos z,\sin z)$ with $f$ of the form
\eqref{decref}, with principal monomial $z^2 v$. One may therefore choose
$\epsilon=\pi /2$, and obtain the necessary and sufficient condition
that $G(y)$ should have exactly $4k+2$ real zeros in 
any interval $I_k=[-2k\pi+\pi/2,2k\pi+\pi/2]$. The condition $G(y)=0$
consists of two branches,
\be
\cos(y/2)=0 \quad \mbox{or} \quad
\tan(y/2) = \frac{2(e_1-e_0)y}{y^2+4 e_0 e_1}
\ee
The first branch gives $2k$ zeros in the interval $I_k$, leaving $2k+2$
zeros to be found on the second branch. Here one meets the argument in
the main text, and find that for $D=e_1-e_0- e_0 e_1 < 0$, there are
$4k$ zeros for $G(k)$ in $I_k$ ($k$ large enough):
this region is therefore unstable. For $D>0 $ and
$e_1- e_0 <-2$,  $G(k)$ has only $4k-2$ zeros
in $I_k$, showing also unstability. 
If $D>0$ and $e_1 - e_0> -2$ on the contrary, $G$ has $4k+2$ zeros
in $I_k$, hence has only real zeros by theorem 2. The motion is therefore
stable.

\section{Open-closed duality} 
\label{opcldual}

In this section, we provide some details on the proof of the
open/closed duality formula \eqref{ocd}, taking inspiration
from a similar discussion in \cite{Gaberdiel:2002hh}.
Let's define $\cal Z$ by:
\be {\cal Z}_{c,o} (t,e_0, e_1) = Z^{cl, op}(t, e_0, e_1)
Z^{cl, op}(t, -e_0, -e_1) = Z^{cl, op}_X (t, e_0, e_1)
Z^{cl, op}_Y (t, e_0, e_1)\ee

We start with the closed string one-loop partition function. Its
logarithm can be written as
\begin{multline}
\log {\cal Z}_c (\hat{t},\hat{e}_0, \hat{e}_1) =  (-2\pi \hat{t}).2
 \underbrace{\left(-\frac1{24}\right)}_{E_0} + 2\log ({\cal N}(\hat{e}_0){\cal N}(\hat{e}_1))
 - \frac12 \log \frac12 \left[\hat{t} + 
i \left(\frac1{\hat{e}_1} -  \frac1{\hat{e}_0}\right) \right] \\
 - \sum_{n=1}^\infty 
\log \left(1-\frac{i\pi n+\hat{e}_0}{i\pi n-\hat{e}_0} \frac{i\pi n-\hat{e}_1}{i\pi n+\hat{e}_1}
 \hat{q}^{n}\right)
+ (\hat{e}_0 \to -\hat{e}_0, \hat{e}_1 \to -\hat{e}_1).
\end{multline}
where $\hat{q} = e^{-2\pi \hat{t}}$.

Our aim is to perform a Poisson resummation on $n$.
For this, it is convenient to add a regularizing mass $\hat{x}$
so as to let the sum on $n$ runs from $-\infty$ to $\infty$:
\begin{multline}
- \sum_{n=1}^\infty \log \left(1-\frac{i\pi n+\hat{e}_0}{i\pi n-\hat{e}_0}
 \frac{{i}\pi n-\hat{e}_1}{{i}\pi n+\hat{e}_1} \hat{q}^{n}\right)\\
= - \frac12 \lim_{x\to 0^+}
 \Biggl[\sum_{n=-\infty}^\infty \log 
\left(1-\frac{i\pi \hat{\chi}_n+\hat{e}_0}{i\pi \hat{\chi}_n-\hat{e}_0}
 \frac{{i}\pi \hat{\chi}_n-\hat{e}_1}{i\pi \hat{\chi}_n+\hat{e}_1}
 \hat{q}^{\hat{\chi}_n}\right) \Biggr. \\
- \Biggl. \log \hat{x}  - \log \left( 2\pi \left(\hat{t}+{i}
 \left( \frac1{\hat{e}_1} - \frac1{\hat{e}_0} \right)\right) \right)
\Biggr]
\end{multline}
where $\hat{\chi}_n = \sqrt{n^2 + \hat{x}^2}$. We now expand out the logarithm
in the sum into
\begin{multline}
S_1 \overset{def}{=} -\frac12 \sum_{n=-\infty}^\infty 
\log \left(1-\frac{{i}\pi \hat{\chi}_n+\hat{e}_0}{i\pi \hat{\chi}_n-\hat{e}_0}
 \frac{i\pi \hat{\chi}_n-\hat{e}_1}{i\pi \hat{\chi}_n+\hat{e}_1} \hat{q}^{\hat{\chi}_n}\right) \\
= \frac12 \sum_{n,p} \left[\left( \frac{i\pi \hat{\chi}_n+\hat{e}_0}{i\pi \hat{\chi}_n-\hat{e}_0}
\frac{i\pi \hat{\chi}_n-\hat{e}_1}{i\pi \hat{\chi}_n+\hat{e}_1}\right)^p + 
 \left( \frac{i\pi \hat{\chi}_n-\hat{e}_0}{i\pi \hat{\chi}_n+\hat{e}_0}
\frac{i\pi \hat{\chi}_n+\hat{e}_1}{i\pi \hat{\chi}_n-\hat{e}_1}\right)^p \right] \frac1p
 \hat{q}^{\hat{\chi}_n p}
\end{multline}
The term $\hat{q}^{\hat{\chi}_n p}/p$ term is best rewritten in integral form 
using a Schwinger time $s$, whereas
the other factor can be expanded in power series with 
respect to $\frac{\hat{\chi}^{2r}}{\hat{e}_0^{r'} \hat{e}_1^{2r-r'}}$. The sum is now
\be
S_1 = \frac1{2\sqrt{\pi}} \sum_{n,p, r, r'} c_{r', 2r-r'}^p
\frac{\hat{\chi}_n^{2r}}{\hat{e}_0^{r'} \hat{e}_1^{2r-r'}} \int_0^\infty \frac{d s}{s^{1/2}}
 e^{-p^2 s -\pi^2 \hat{t}^2 (n^2 + x^2)/s}
\ee
$\hat{\chi}_n^{2r}$ terms can be generated by differentiating the integral on $s$
with respect to $\hat{t}^2$, thus yielding an expression which can be 
Poisson resummed at ease:
\begin{align}
S_1 &= \frac1{2\sqrt{\pi}} \sum_{n,p, r, r'}
\frac{c_{r', 2r-r'}^p}
{\hat{e}_0^{r'} \hat{e}_1^{2r-r'}} \frac{\partial^r}{(\partial \hat{t}^2)^r} 
\int_0^\infty \frac{d s}{s^{1/2}} \left(\frac{-s}{\pi^2} \right)^r
 e^{-p^2 s -\pi^2 \hat{t}^2 (n^2+x^2)/s} \\
  &= \frac1{2\pi} \sum_{\hat{n},p, r, r'} \frac{c_{r', 2r-r'}^p}{\hat{e}_0^{r'} \hat{e}_1^{2r-r'}}
 \frac{\partial^r}{(\partial \hat{t}^2)^r}
\int_0^\infty d s \frac1t  \left(\frac{-s}{\pi^2} \right)^r
 e^{-p^2 s - \hat{n}^2 s/t^2 - \pi^2 \hat{t}^2 x^2/s}
\end{align}
The sum of $\hat{n}$ over $\mathbb{Z}$ can be folded into  a sum 
over $\mathbb{N}$ at the expense of a factor of 2.
Let us change variable $\tilde s = \frac{s}{\hat{t}^2}$, 
expand $e^{-\hat{t}^2 p^2 \tilde s}$
and differentiate $r$ times with respect to $\hat{t}^2$ 
in $S'_1$ (the $'$ means: without $\hat{n} =0$ term). We obtain
\be S'_1= \frac1\pi \sum_{\hat n, p, r, r'} \int_0^\infty \hspace{-.2cm} d \tilde s \sum_{l=0}^\infty
\frac{(-p^2 \tilde s)^l}{l!}
\frac{\Gamma(l+r+\frac32)}{\Gamma(l+\frac32)} \frac{c_{r', 2r-r'}^p}{\hat{e}_0^{r'} \hat{e}_1^{2r-r'}} 
\left(\frac{(-\tilde s)}{\pi^2}\right)^r \hat{t}^{2l+1}
e^{-\hat{n}^2 \tilde s - \pi^2 x^2/\tilde s} \ee
and for the $\hat{n} =0$ term,
\be T_0 = \frac1{2\pi} \sum_{p,r,r'} c_{r', 2r-r'}^p
\frac{\partial^r}{(\partial \hat{t}^2)^r} 
\int_0^\infty d \tilde s \, \hat{t}^2 \frac{(-\tilde s \hat{t}^2)^r}{\pi^{2r} \hat{e}_0^{r'}
\hat{e}_1^{2r-r'}}
\frac1{\hat{t}} \epn^{-p^2 \tilde s \hat{t}^2} \ee
Gathering terms, one gets
\be S'_1 = \log {\cal Z}_c + 4\pi \hat{t} E_0 - \log (4\pi x) - T_0 
- 2\log ({\cal N}(\hat{e}_0){\cal N}(\hat{e}_1)) \ee

Let us now turn to the open channel.
The logarithm of the open string partition function is
\be \log {\cal Z}_o (t,e_0, e_1) =  -(2\pi t) (E_X + E_Y)
 - \sum_{p=0}^\infty \log \left(1-q^{\om_p^X}\right)
 - \sum_{p=0}^\infty \log \left(1-q^{\om_p^Y}\right)
\ee
where $q = e^{-2 \pi t}$ and assuming that $\om_0^{X,Y} \neq 0$.
We denote by $P_{X,Y}$ the set of solutions (renamed $\hat{p}$) to the 
open string dispersion relation \emph{including} 
$\om=0$ solution. Introducing a regularizing mass $x$, the two sums 
can be rewritten (here we give only the expressions for $X$)
\begin{multline}
-\sum_{p=0}^\infty \log \left(1-q^{\om_p^X}\right) =
-\lim_{x \to 0^+} \sum_{p=0}^\infty \log \left(1-q^{\sqrt{\left(\om_p^X\right)^2 + x^2}}\right) = \\
= -\frac12 \lim_{x \to 0^+} \sum_{\hat{p} \in P_X} \log \left(1-q^{\chi_{\hat{p}}}\right) 
+ \frac12 \lim_{x \to 0^+} \log \left(1-q^x\right).
\end{multline}
Note that we made a difference between the regularizing masses of closed and open strings because, as
the duality involves a conformal factor $\frac1t$ between the annulus and the 
cylinder, hence $ x = \frac{\hat{x}}t$. Therefore
\be \log \left(1-q^x\right) \underset{x\to 0}{=} \log \hat{x} + \log (2\pi) +{\cal O}(x).\ee
We now expand the log and represent the power of $q$ by a Schwinger-type
integral, obtaining in the limit $x \to 0$
\be
S_2 \overset{def}{=} 
-\frac12 \sum_{\hat{p} \in P_X, P_Y} \log \left(1-q^{\chi_{\hat{p}}}\right)
 =  \frac1{2\sqrt{\pi}} \sum_{\hat{n} \in \mathbb{N}^*,\: \hat{p}\in P_X,\, P_Y}
  \int_0^\infty \frac{d \tilde s}{\tilde{s}^{1/2}}
e^{-\hat{n}^2 \tilde s - \pi^2 t^2 (\hat{p}^2 +x^2)/ \tilde s}
\ee
We now represent the above sum by a contour integral
\begin{multline}
 \sum_{\hat{p} \in P_X, P_Y} e^{-\pi^2 t^2 \hat{p}^2 /\tilde s}
= \oint_{\cal C} d \rho \, e^{-\pi^2 t^2 \rho^2 /\tilde s}  \\
\times \left[ \frac{1-\dfrac{e_0}{\pi^2 \rho^2+e_0^2} + \dfrac{e_1}{\pi^2 \rho^2+e_1^2}}
{1-\dfrac{{i}\pi\rho + e_0}{{i}\pi\rho - e_0}\dfrac{{i}\pi\rho - e_1}{{i}\pi\rho + e_1}
e^{-2{i}\pi\rho}} + 
\frac{1+\dfrac{e_0}{\pi^2 \rho^2+e_0^2} - \dfrac{e_1}{\pi^2 \rho^2+e_1^2}}
{1-\dfrac{{i}\pi\rho - e_0}{{i}\pi\rho + e_0}\dfrac{{i}\pi\rho + e_1}{{i}\pi\rho - e_1}
e^{-2{i}\pi\rho}}  \right] \label{contourint}
\end{multline} 
where $\cal C$ is a contour encircling all the $\hat{p} \in P_X, P_Y$.
$\cal C$ can be deformed into two lines $L_1$ 
and $L_2$, where $L_1$ runs from $\infty+i\epsilon$ to $-\infty+i\epsilon$ 
above the real axis and $L_2$ runs 
from $-\infty-i\epsilon$ to $\infty-i\epsilon$ below the real axis.
In order to expand the denominators of \eqref{contourint} 
in power series of $\rho^2$, one have to push $L_1$ and $L_2$
to the region of the complex plane where $\left\vert Im(z) \right\vert \ge e_0, e_1$. The integral then picks
contribution from the poles of the numerators 
of \eqref{contourint}, hence giving the following contribution
\begin{multline}
\frac1{\sqrt{\pi}} \sum_{\hat{n}=1}^\infty \int_0^\infty \frac{d \tilde s}{\tilde{s}^{1/2}} \left(
e^{-\hat{n}^2 \tilde s + \pi^2 t^2 \left(\left(\pm\frac{ie_0}{\pi}\right)^2 +x^2\right)/ \tilde s} + 
e^{-\hat{n}^2 \tilde s + \pi^2 t^2 \left(\left(\pm\frac{ie_1}{\pi}\right)^2 +x^2\right)/ \tilde s}
\right) \\
\begin{aligned}
 &= -\frac12\log (1-e^{2\hat{e}_0})(1-e^{-2\hat{e}_0})(1-e^{2\hat{e}_1})(1-e^{-2\hat{e}_1})\\
 &= -\log \left(4 \sinh(\hat{e}_0) \sinh(\hat{e}_1)\right),
\end{aligned}
\end{multline} 
where we anticipate the relation between $e_i$ and $\hat{e}_i$, \eqref{rescaling}.

Expanding the denominators in \eqref{contourint}, one can see 
that the contour integrals over $L_1$ and $L_2$ are equal up to the $p=0$ term
of the power serie expansion.
This term gives the vacuum energy of the closed string computation: it reads
\be \frac1{\sqrt{\pi}} \sum_{\hat{n}=1}^\infty \int_{-\infty}^\infty d \rho
 \int_0^\infty \frac{d \tilde s}{\tilde{s}^{1/2}}
e^{-\hat{n}^2 \tilde s - \pi^2 t^2 (\rho^2 +x^2)/ \tilde s} =
 \frac{-2(2\pi)}t\left(-\frac1{24}\right) = -2\pi \frac1t (2E_0).
\ee
Terms with non vanishing $p$ are on the other hand, summing $L_1$ and $L_2$ equal terms,
\begin{multline}
\frac1{\sqrt{\pi}} \sum_{\hat{n},\, p\in\mathbb{N}^*} \int_{-\infty}^\infty d \rho
\Biggl( \left[ 
\left(\dfrac{{i}\pi\rho + e_0}{{i}\pi\rho - e_0}\dfrac{{i}\pi\rho - e_1}{{i}\pi\rho + e_1}
\right)^p 
+ \left(\dfrac{{i}\pi\rho - e_0}{{i}\pi\rho + e_0}\dfrac{{i}\pi\rho + e_1}{{i}\pi\rho - e_1}
\right)^p \right]\Biggr. \\
\Biggl. - \left(\dfrac{e_0}{\pi^2 \rho^2+e_0^2} - \dfrac{e_1}{\pi^2 \rho^2+e_1^2}\right)
\left[ 
\left(\dfrac{{i}\pi\rho + e_0}{{i}\pi\rho - e_0}\dfrac{{i}\pi\rho - e_1}{{i}\pi\rho + e_1}
\right)^p
- \left(\dfrac{{i}\pi\rho - e_0}{{i}\pi\rho + e_0}\dfrac{{i}\pi\rho + e_1}{{i}\pi\rho - e_1}
\right)^p \right] \Biggr) \\
\times e^{-2{i} \pi \rho p} \int_0^\infty \frac{d \tilde s}{\tilde{s}^{1/2}}
e^{-\hat{n}^2 \tilde s - \pi^2 t^2 (\hat{p}^2 +x^2)/ \tilde s}.
\end{multline}
Changing variable to
\be e^{2{i} \phi} =
\dfrac{{i}\pi\rho + e_0}{{i}\pi\rho - e_0}\dfrac{{i}\pi\rho - e_1}{{i}\pi\rho + e_1},
\ee 
one finds the following expression,
\be  S'_2 = \frac1{\pi\sqrt{\pi}}
\int d \rho \sum_{\hat{n},p} 
\left( 2\pi \cos 2p\phi + \frac{{i}}p \frac{d \cos 2p\phi}{d \rho} \right) e^{-2{i} \pi \rho p}
\int_0^\infty \frac{d \tilde s}{\tilde{s}^{1/2}}
e^{-\hat{n}^2 \tilde s - \pi^2 t^2 (\hat{p}^2 +x^2)/ \tilde s},
\ee
where $S'_2$ is the sum without $p=0$ term.
Integrating by part on $\rho$ the second term, one gets 
\be
S'_2 = -\frac{i}{\pi\sqrt{\pi}} \sum_{\hat{n}, p} \int_0^\infty \frac{d \tilde s}{\tilde{s}^{1/2}} 
\int_{-\infty}^\infty d \rho\, e^{-2{i} \pi \rho p} \frac1p  \cos 2p\phi \,
\frac{d}{d \rho} e^{-\hat{n}^2 \tilde s - \pi^2 t^2 (\rho^2 +x^2)/ \tilde s} .
\ee
Let us now differentiate with respect to $\rho$, 
then expand $2 \cos(2 p \phi)$ (with the \emph{same} coefficients
as for the open string computation) and convert $\rho^{2r}$ term 
to $\frac{\partial^r}{(\partial t^2)^r}$
and the additional $\rho$ factor to $\frac{\partial}{\partial p}$
\begin{multline}
S'_2 = \frac1{2\sqrt{\pi}} \sum_{\hat{n}, p, r, r'} \int_0^\infty \hspace{-.2cm} \frac{d\tilde s}{\tilde{s}^{3/2}} 
\int_{-\infty}^\infty d \rho  \\ \times
\frac{t^2}p \frac{c_{r', 2r-r'}^p}{e_0^{r'} e_1^{2r-r'}} \left(\frac{-\tilde s}{\pi^2} \right)^r
e^{-\hat{n}^2 \tilde s - \pi^2 t^2 x^2/ \tilde s} \frac{\partial^r}{(\partial t^2)^r}
\frac{\partial}{\partial p}\, \frac1{t}\, e^{-2{i}\pi\rho p} e^{- \pi^2 t^2 \rho^2/ \tilde s}
\end{multline}
Now we can integrate over $\rho$ and differentiate with respect to $p$,
\be
S'_2 = \frac1{\pi} \sum_{\hat n =1}^\infty \int_0^\infty d\tilde s \sum_{p=1}^\infty  
\sum_{r,r'} 
c_{r', 2r-r'}^p \frac1{e_0^{r'} e_1^{2r-r'}} \left(\frac{-\tilde s}{\pi^2} \right)^r t^2
\, e^{-\hat{n}^2 \tilde s - \pi^2 t^2 x^2/ \tilde s}   
\frac{\partial^r}{(\partial t^2)^r} \frac1{t^3} \, e^{-p^2 \tilde{s}/t^2}
\ee
Expanding the exponential in power series to differentiate with respect to $t^2$, one finds
\be
S'_2 = \frac1{\pi} \sum_{\hat n, p, r, r', l}  \int_0^\infty d\tilde s   
\frac{c_{r', 2r-r'}^p}{e_0^{r'} e_1^{2r-r'}} \left(\frac{+\tilde s}{\pi^2} \right)^r \,
e^{-\hat{n}^2 \tilde s - \pi^2 t^2 x^2/ \tilde s} \frac{(-p^2 \tilde{s})^l}{l!} 
\frac{\Gamma(l+r+\tfrac32)}{\Gamma(l+\tfrac32)} t^{-2l-2r-1}, 
\ee
and thus conditions of equality between $S'_1$ (closed side) and $S'_2$ (open side)
\be
t = \dfrac1{\hat{t}}\ ,\quad e_i = i \dfrac{\hat{e}_i}{t} = i \hat{t} \hat{e}_i \label{rescaling}
\ee
There remains the vacuum energy term. We show that it is equal to $T_0$ (substracting divergences).
\be
\frac1{2\sqrt{\pi}} \sum_{\hat p} \int_0^\infty  \frac{d \tilde s}{\tilde{s}^{1/2}}\, 
e^{-\hat{n}^2 \tilde s - \pi^2 t^2 (\hat{p}^2 +x^2)/ \tilde s} 
= \frac12 \sum_{\hat p} \frac1{\hat n} e^{-2\pi \hat n t \sqrt{\hat{p}^2 +x^2}}
\underset{\hat{n} \to 0}{=} \frac12 \sum_{\hat p} \left( \frac1{\hat n} 
- 2\pi t \lvert\hat p\rvert\right)
\ee
where $\lim_{x \to 0}$ is assumed in the last line. On the other hand, 
the vacuum energy reads
\be
E_X + E_Y = \frac12 \sum_{p=0}^\infty (\om^X_p + \om_p^Y) 
= \frac14 \sum_{\hat p \in P_X,\, P_Y} \lvert\hat p \rvert \ee
Thus vacuum energy is related to the term $\hat n =0$ (substracting divergences) 
in the general computation. 
Then, performing the same manipulation than on the general $\hat{n}$ terms, and removing 
the linear and quadratic divergences arising in the definition of the 
vacuum energy, one can show that
\be
 -(2\pi t)(E_X+E_Y) = \text{(term }\hat{n} = 0 \text{ in the closed side)},
\ee
Then
\be S'_2 = \log {\cal Z}_o  
+ \log\left(\frac{4 \sinh(\hat{e}_0) \sinh(\hat{e}_1)}{2\pi \hat{x}}\right)
+  \frac{4\pi E_0}t - T_0
\ee

From $S'_1 = S'_2$, one gets the following duality formula
\be {\cal Z}_c (\hat{t}, \hat{e}_0, \hat{e}_1) = ({\cal N}(\hat{e}_0))^2({\cal N}(\hat{e}_1))^2
\left(8 \sinh(\hat{e}_0) \sinh(\hat{e}_1)\right)
{\cal Z}_o (\frac1{\hat{t}}, i \hat{e}_0 \hat{t}, i \hat{e}_1 \hat{t})  \label{final}\ee
Normalisation of the boundary states
\be {\cal N}(\hat{e}) = \frac1{\sqrt{2\sqrt{2} \sinh(\hat{e})}}. \label{normbndry} \ee

\section{Normalization of eigenmodes}
\label{norm}
\subsection{Without magnetic field}
\begin{itemize}
\item Normalization of the excited modes ($n \neq 0$) and of mode 0 when $D>0$,
\be
\begin{split}
{\cal N}_n  &= \left(4\pi + 
\frac{i}{\om_n} \left(\frac{1}{\Omega_n^1} - \frac{1}{\Omega_n^0} - (\Omega_n^1 - \Omega_n^0)\right)\right)^{-\frac12} \\
     &= \left[4\pi \left(1 + \left(\frac{e_1}{(\pi\om_n)^2+e_1^2} - \frac{e_0}{(\pi\om_n)^2+e_0^2}\right)\right)\right]^{-\frac12},
\end{split}
\ee
with $\Omega_n^i = \frac{i\pi \om_n - e_i}{i\pi \om_n +e_i}$,

\item Normalization of the zero mode when $D=0$, 
\be {\cal N}_0^c = \left(\frac{3}{\pi(3(1-e_0)+e_0^2)}\right)^{\frac12},
\ee
\item Normalization of the zero mode when $D<0$,
\be {\cal N}^u = \left[4\pi\left( 1+ \left(\frac{e_1}{-(\pi k_0)^2+e_1^2} - \frac{e_0}{-(\pi k_0)^2+e_0^2}\right)\right)\right]^{-\frac12},
\ee
(Note that ${\cal N}^u = {\cal N}_0$ replacing $\om_0$ by $ik_0$.)
\item Normalization of the zero mode when $D<0$ and $k_0 = \pm \frac{e}{\pi}$,
\be {\cal N}^u_e = \left(\frac{e}{\pi \left(1- e^{-2 e} \right)}\right)^{\frac12}.
\ee
\end{itemize}

With the normalization chosen, we have the following orthogonality
property between the $Y_n$ defined
by
\be Y_n = {\cal N}_n \frac{i}{\omega_n} \left( 
e^{-i\om_n(\tau+\sigma)} + 
\frac{i\pi\om_n-e_0}{i\pi\om_n+e_0} e^{-i\om_n (\tau-\sigma)} \right)\ , \ee
\be \int_0^\pi \left(Y_n \pt Y_m - Y_m \pt Y_n \right) d\sigma = 
\frac{i}{\om_n}\delta_{n,-m} \ .
\label{orthogonality}
\ee

\subsection{With magnetic field}

The normalisation coefficient then become matrices $\cal M$. We quote 
only the coefficients 
needed for the mode expansion \eqref{modexpeq}.

\begin{itemize}
\item Normalization of the excited modes ($n \neq 0$),
\be
{\cal M}_n  =  \frac1{\sqrt{4\pi}},
\ee

\item Normalization of the zero mode for the two unstable modes: $k_0 = \pm \frac{e}{1\pm B}$,
\be \left({\cal M}^u_+\right)^T{\cal M}^u_- =
\left[ \frac{e}{\pi \left(1-e^{-2\frac{e}{1+B}}\right)} \right] .
\ee

\end{itemize}

\end{document}